\documentclass[aps, prx, twocolumn, letterpaper]{revtex4-2}

\usepackage{graphicx}
\usepackage{amssymb}
\usepackage{amsmath}
\usepackage{float}
\usepackage{txfonts}
\usepackage{bm}
\usepackage{color}
\usepackage[colorlinks=true, linkcolor=blue, anchorcolor=black, citecolor=blue, filecolor=black, menucolor=black, runcolor=black, urlcolor=blue]{hyperref}
\usepackage{mathrsfs}
\usepackage{multirow}

\newcommand{\Z}{$n+1$}
\newcommand{\Zt}{$2$}
\newcommand{\ZZ}{$\left(\frac{n}{2}+1\right)^2$}
\newcommand{\Zd}{$\frac{n}{2}+1$}
\newcommand{\Ztt}{$2\times 2$}
\newcommand{\Zdd}{$\left(\frac{n}{4}+1\right)^2$}
\newcommand{\Lr}{~${\rm L}_{\rm r}$}
\newcommand{\Li}{~${\rm L}_{\rm i}$}

\newcommand{\Ca}{ \Z & $1$ & \Z & $1$ & \Z & $1$ & \Z & $1$ & \Z & $1$}
\newcommand{\Caa}{ \ZZ & $1$ & \ZZ & $1$ & \ZZ & $1$ & \ZZ & $1$ & \ZZ & $1$}
\newcommand{\Cb}{ $1$ & \Z & $1$ & \Z & $1$ & \Z & $1$ & \Z & $1$ & \Z}
\newcommand{\Cbb}{ $1$ & \ZZ  & $1$ & \ZZ  & $1$ & \ZZ  & $1$ & \ZZ & $1$ & \ZZ }

\newcommand{\Ra}{\Z & $1$ & $1$ & $1$ & \Zd & $1$ & \Zt & \Zt & \Z & $1$}
\newcommand{\Raa}{ \ZZ & $1$ & $1$ & $1$ & \Zdd & $1$ & \Ztt & \Ztt & \ZZ & $1$}

\newcommand{\Rb}{ \Zt & \Z & $1$ & $1$ & $1$ & \Zd & $1$ & \Zt & \Zt & \Z}
\newcommand{\Rbb}{ \Ztt & \ZZ & $1$ & $1$ & $1$ & \Zd & $1$ & \Ztt &  \Ztt & \ZZ}

\newcommand{\Rc}{\Zt & \Zt & \Z & $1$ & $1$ & $1$ & \Zd & $1$ & \Zt & \Zt}
\newcommand{\Rcc}{ \Ztt & \Ztt & \ZZ & $1$ & $1$ & $1$ & \Zdd & $1$ & \Ztt & \Ztt }

\newcommand{\Rd}{$1$ & \Zt & \Zt & \Z & $1$ & $1$ & $1$ & \Zd & $1$ & \Zt}
\newcommand{\Rdd}{ $1$ & \Ztt & \Ztt & \ZZ & $1$ & $1$ & $1$ & \Zdd & $1$ & \Ztt}

\newcommand{\Ree}{\Zd & $1$ & \Zt & \Zt & \Z & $1$ & $1$ & $1$ & \Zd & $1$}
\newcommand{\Reee}{\Zdd & $1$ & \Ztt & \Ztt & \ZZ & $1$ & $1$ & $1$ & \Zdd & $1$}

\newcommand{\Rf}{ $1$ & \Zd & $1$ & \Zt & \Zt & \Z & $1$ & $1$ & $1$ & \Zd}
\newcommand{\Rff}{ $1$ & \Zdd & $1$ & \Ztt & \Ztt & \ZZ & $1$ & $1$ & $1$ & \Zdd}

\newcommand{\Rg}{ $1$ & $1$ & \Zd & $1$ & \Zt & \Zt & \Z & $1$ & $1$ & $1$}
\newcommand{\Rgg}{ $1$ & $1$ & \Zdd & $1$ & \Ztt & \Ztt & \ZZ & $1$ & $1$ & $1$}

\newcommand{\Rh}{ $1$ & $1$ & $1$ & \Zd & $1$ & \Zt & \Zt & \Z & $1$ & $1$}
\newcommand{\Rhh}{ $1$ & $1$ & $1$ & \Zdd & $1$ & \Ztt & \Ztt & \ZZ & $1$ & $1$}

\newcommand{\PCa}{ \Z & $1$ & \Z & $1$ & \Z & $1$ & \Z & $1$ & \Z & $1$}

\newcommand{\PCb}{  $1$ & \Z & $1$ & \Z & $1$ & \Z & $1$ & \Z & $1$ & \Z}

\newcommand{\PRa}{\Z &  \Zt & \Zt & $1$ & \Zd & $1$ & $1$ & $1$ & \Z &  \Zt}
\newcommand{\PRaa}{\ZZ &  \Ztt & \Ztt & $1$ & \Zdd & $1$ & $1$ & $1$ & \ZZ &  \Ztt}

\newcommand{\PRb}{ \Zt & \Zt & $1$ & \Zd & $1$ & $1$ & $1$ & \Z  & \Zt &  \Zt }
\newcommand{\PRbb}{ \Ztt & \Ztt & $1$ & \Zdd & $1$ & $1$ & $1$ & \ZZ  & \Ztt & \Ztt }

\newcommand{\PRc}{\Zt & $1$ & \Zd & $1$ & $1$ & $1$ & \Z & \Zt & \Zt & $1$}
\newcommand{\PRcc}{\Ztt & $1$ & \Zdd & $1$ & $1$ & $1$ & \ZZ & \Ztt & \Ztt & $1$}

\newcommand{\PRd}{$1$ & \Zd & $1$ & $1$ & $1$ & \Z & \Zt & \Zt & $1$ & \Zd}
\newcommand{\PRdd}{ $1$ & \Zdd & $1$ & $1$ & $1$ & \ZZ & \Ztt & \Ztt & $1$ & \Zdd}

\newcommand{\PRee}{\Zd & $1$ & $1$ & $1$ & \Z & \Zt &  \Zt & $1$ & \Zd & $1$}
\newcommand{\PReee}{\Zdd & $1$ & $1$ & $1$ & \ZZ & \Ztt &  \Ztt & $1$ & \Zdd & $1$}

\newcommand{\PRf}{$1$ & $1$ & $1$ & \Z & \Zt & \Zt & $1$ & \Zd & $1$ & $1$}
\newcommand{\PRff}{$1$ & $1$ & $1$ & \ZZ & \Ztt & \Ztt & $1$ & \Zdd & $1$ & $1$}

\newcommand{\PRg}{$1$ & $1$ & \Z & \Zt & \Zt & $1$ & \Zd & $1$ & $1$ & $1$}
\newcommand{\PRgg}{$1$ & $1$ & \ZZ & \Ztt & \Ztt & $1$ & \Zdd & $1$ & $1$ & $1$}

\newcommand{\PRh}{ $1$ & \Z & \Zt & \Zt & $1$ & \Zd & $1$ & $1$ & $1$ & \Z}
\newcommand{\PRhh}{ $1$ & \ZZ & \Ztt & \Ztt & $1$ & \Zdd & $1$ & $1$ & $1$ & \ZZ}

\begin{document}


\preprint{APS/123-QED}

\title{Unsupervised Classification of Non-Hermitian Topological Phases under Symmetries}

\author{Yang Long$^{1,2}$}
\email{longyangphysics@tongji.edu.cn}
\author{Haoran Xue$^3$}
\author{Baile Zhang$^{2,4}$}
\email{blzhang@ntu.edu.sg}
\affiliation{%
$^1$School of Physics Science and Engineering, Tongji University, Shanghai 200092, China \\
$^2$Division of Physics and Applied Physics, School of Physical and Mathematical Sciences, Nanyang Technological University, 21 Nanyang Link, Singapore 637371, Singapore \\
$^3$Department of Physics, The Chinese University of Hong Kong, Shatin, Hong Kong SAR, China \\
$^4$Centre for Disruptive Photonic Technologies, Nanyang Technological University, Singapore 637371, Singapore
}%

\begin{abstract}
The integration of artificial intelligence (AI) into fundamental science has opened new possibilities to address long-standing scientific challenges rooted in mathematical limitations. For example, topological invariants are used to characterize topology, but there is no universally applicable one. This limitation explains why, in the past decades-long classification of topological phases of matter---mainly focused on Hermitian systems---many phases initially classified ``trivial" were later identified as topological. Recently, the discovery of non-Hermitian band topology has spurred substantial efforts in non-Hermitian topological classification, including the development of new topological invariants. However, such classifications similarly risk overlooking key topological features.
Here, without relying on any topological invariant, we develop an AI-based unsupervised classification of symmetry-protected non-Hermitian topological phases. This algorithm distinguishes topological differences among non-Hermitian Hamiltonians with symmetries, and constructs, in an unsupervised manner, a topological periodic table for non-Hermitian systems. Additionally, it can account for the boundary effects, enabling the exploration of open-boundary effects on the topological phase diagram.       
These results introduce an unsupervised approach for classifying symmetry-protected non-Hermitian topological phases without omission and provide valuable guidance for the development of theories and experiments. 
\end{abstract}

\maketitle

Artificial intelligence (AI) for science, often referred to as ``AI for Science", leverages human-like intelligence to handle fundamental scientific problems~\cite{Wang2023, Krenn2022,Mehta2019, Dunjko2018, Carleo2019}.  
For example, AI techniques commonly used in computer vision have been employed to distinguish paramagnetic and ferromagnetic phases in the Ising model~\cite{Nieuwenburg2017, Carrasquilla2017}.
The restricted Boltzmann machines, traditionally used for dimensionality reduction~\cite{Hinton2006}, can efficiently obtain the ground states of many-body systems~\cite{Carleo2017, Melko2019}.  
To date, AI has demonstrated its capability to solve problems either at a human level~\cite{Mnih2015} or with significantly higher efficiency in game playing and protein predictions~\cite{Silver2016, Silver2017, Silver2018,Senior2020, Jumper2021}. However, most applications are still constrained by human knowledge, addressing challenges within existing mathematical frameworks rather than surpassing the mathematical limitations that underpin many branches of physics.

Topology is a mathematical concept describing properties of objects preserved during continuous deformation. Its use to characterizing topological phases of matter has revolutionized condensed matter physics in the past decades~\cite{Bansil2016, Haldane2017, Hasan2010, Qi2011}. 
A long-standing challenge in topology, which has carried over to the classification of topological phases of matter, is the lack of a universally applicable topological invariant. This limitation leads to a fundamental risk: even if all existing topological invariants identify a phase (or a specific Hamiltonian) as ``trivial", it may later be identified as topological with the development of new invariants. This issue arose in earlier classification of Hermitian topological phases, where phases initially deemed ``trivial", such as the topological valley Hall phase~\cite{Zhang2013} and higher-order topological phases~\cite{Benalcazar2017, Liu2017}, were subsequently recognized as topological through theoretical advances.

This challenge is particularly pronounced in the emerging field of non-Hermitian topological phases~\cite{Moiseyev2011, Bender2007,  Li2023a, Feng2017, ElGanainy2018, Bergholtz_2021}. Unlike Hermitian topological phases, which have accumulated numerous topological invariants through decades of extensive studies~\cite{Chiu_2016}, the framework of non-Hermitian topological classification is relatively new and remains under active exploration. Significant efforts have been made in recent years to develop new topological invariants to characterize the non-Hermitian topology~\cite{Gong_2018,Kawabata_2019,Zhou_2019,Wojcik_2020}. At the same time, there is a considerable interest in exploring non-Hermitian topological phases~\cite{Miri2019, Coulais2020, Hu2021, Wang2022, Borgnia2020}, due to their promising applications, including high-precision sensors~\cite{Chen2017, Hodaei2017}, mode switching~\cite{Doppler2016, Nasari2022} and high-quality lasers~\cite{Hodaei2014, Feng2014, Zhu2022}.

Unsupervised learning, a major branch of AI, uncovers hidden patterns in raw data without requiring labelled training sets. 
It has proven effective in recognizing topological phases without relying on predefined topological invariants~\cite{Rodriguez_Nieva_2019, Wang2016, Wetzel2017, Scheurer_2020, Che2020, Balabanov2020, Park2022, Ma2022}, addressing challenges such as randomness and disorder that extend beyond theoretical limitations~\cite{Long2020}. 
Recently, unsupervised learning has been successfully applied in the topological classification of Hermitian systems under symmetry constraints~\cite{Long_2023}, generating the topological periodic table in a data-driven manner — an achievement previously possible only through abstract group theory. 
This capability positions unsupervised learning as a promising tool for advancing topological classifications of non-Hermitian systems, overcoming the fundamental limitations of traditional methods~\cite{Yu_2021, Li2023}. 
Early studies have demonstrated its effectiveness in capturing the braiding and knot topology in non-Hermitian bands~\cite{Long2024, Chen2024, Yu2022}. 

In this work, we demonstrate the use of unsupervised learning in topological classification of non-Hermitian systems under symmetries. 
Guided by the fundamental principles of non-Hermitian topology, we introduce a similarity function to identify topological differences based on three distinct gap types: point gap, real line gap, and imaginary line gap. 
Using an unsupervised clustering algorithm~\cite{Long_2023}, we determine the number of phases in Hamiltonian samples and classify each phase unsupervisedly. 
We validate our algorithm through multiple examples, focusing on topological phases induced by non-Hermiticity.  Applying our algorithm to random Hamiltonian samples from non-Hermitian 38-fold symmetry classes, we construct a topological periodic table for non-Hermitian systems. 
This table aligns with theoretical predictions derived from homotopy groups and Clifford algebra~\cite{Gong_2018,Kawabata_2019,Zhou_2019}, which primarily address abstract Hamiltonians but do not directly apply to specific Hamiltonians derived from concrete physical systems.  
In contrast, our algorithm can handle both concrete non-Hermitian systems and random Hamiltonians under symmetries. 
Furthermore, we discuss the effect of parity transformation on symmetry classes, leading to a new topological periodic table for symmetry classes that include parity transformation. 
Finally, we investigate the boundary effects on the non-Hermitian topological phase diagram under open boundary conditions.  
Compared to our previous work on topological classifications for Hermitian systems~\cite{Long_2023}, we here discuss non-Hermitian systems, extend our algorithms to work on new types of band topology on complex-energy plane that are absent in Hermitian systems, and investigate the effect of non-Hermiticity-enriched symmetry classes on topological classifications.

\begin{figure}[tp!]
\centering
\includegraphics[width=\linewidth]{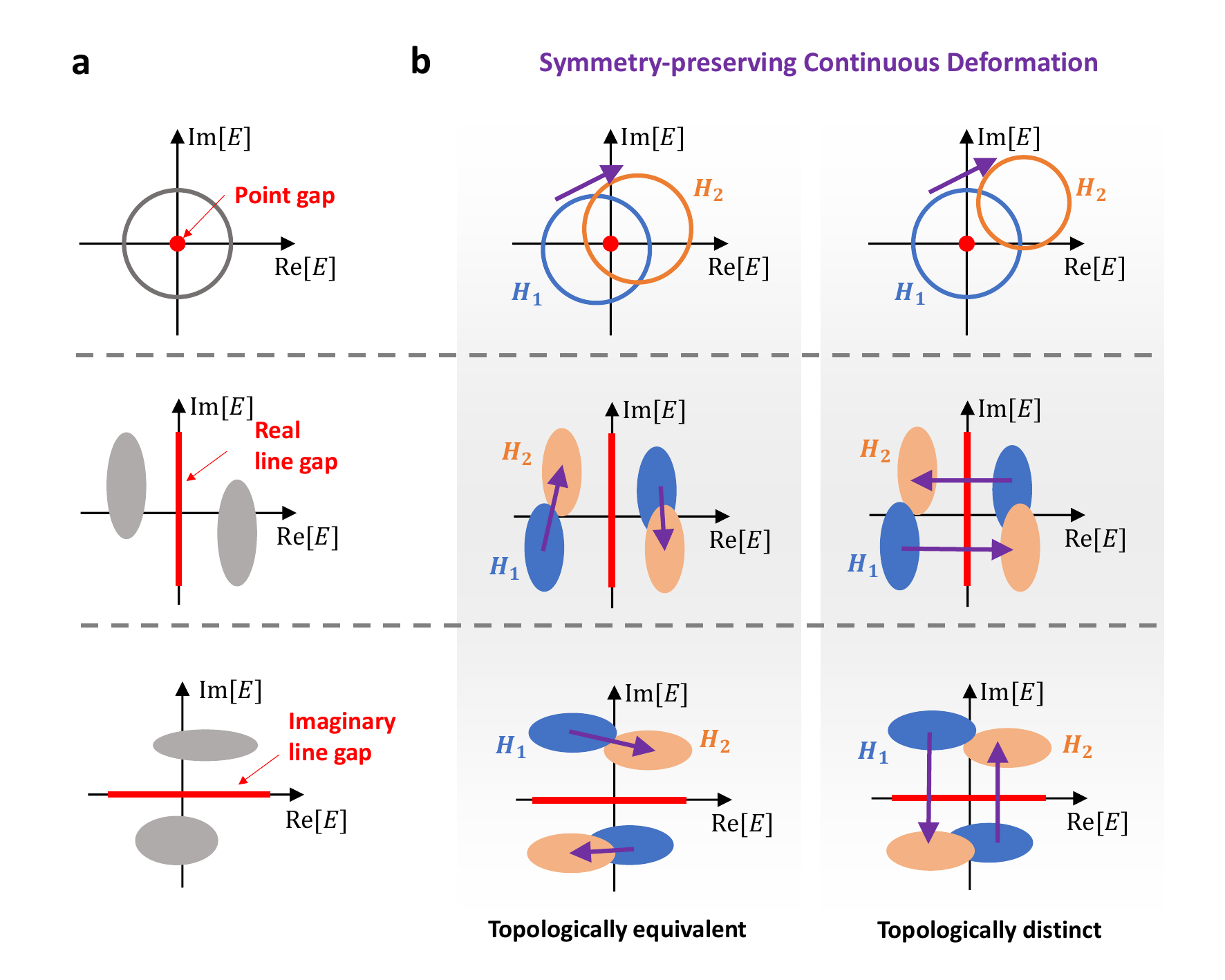}
\caption{Gap types in non-Hermitian systems and symmetry-preserving continuous deformation. (a) Typical three gap types : point gap, real line gap, and imaginary line gap. The gray regions denote the regions covered by the eigen-energies of the Hamiltonian on the complex-energy plane. (b) Symmetry-preserving continuous deformation between two Hamiltonians $H_1$ and $H_2$. When $H_1$ and $H_2$ are topologically equivalent, one can find a path to realize the continuous deformation between them without closing the gap. While $H_1$ and $H_2$ are topologically distinct, any continuous deformation between them will close the gap. The purple arrows denote the continuous deformation. }
\label{fig:mechanism}
\end{figure}

Hermiticity, expressed as $H=H^\dagger$, acts as a symmetry condition that
constrains the allowed terms in the Hamiltonian and protects the real nature of eigen-energies, similar to the continuous time-translation symmetry in Noether's theorem~\cite{Noether1918}. 
Breaking Hermiticity (i.e., $H\neq H^\dagger$) primarily leads to the emergence of complex-energy spectra, but also introduce new topological phases. 
One notable consequence of breaking Hermiticity is the diversification of energy gap types. 
While Hermitian systems typically exhibit only one type of gap, non-Hermitian systems can manifest three distinct types of gaps~\cite{Kawabata_2019, Liu2023}: point gap, real line gap, and imaginary line gap, as illustrated in Fig.~\ref{fig:mechanism}(a). 
A complex energy $E_f \in \mathbb{C}$, often referred as the ``Fermi level", serves as a reference point, with the gaps defined as: (1) a point gap at the specific energy $E_f$; (2) a real line gap along the line defined by ${\rm Re}[E_f]$; and (3) an imaginary line gap along the line defined by ${\rm Im}[E_f]$. 
Furthermore, breaking Hermiticity modifies the symmetry conditions.  
For example, in Hermitian systems, the chiral symmetry $\Gamma$ ($U_{\Gamma} H^\dagger(\bm{k}) U_{\Gamma}^{-1} = -H(\bm{k}) $) and the sublattice symmetry $\mathcal{S}$ ($U_{\mathcal{S}}  H(\bm{k}) U_{\mathcal{S}}^{-1} = -H(\bm{k}) $) are identical. In contrast, for non-Hermitian systems, the chiral symmetry and sublattice symmetry are different, leading to a richer variety of symmetries and topological phases. 

Intuitively, similar to Hermitian systems, the topological phases of non-Hermitian systems are  defined by whether there exists a continuous deformation path between two Hamiltonians without closing the gap. 
Specifically, as illustrated in Fig.~\ref{fig:mechanism}(b), if two non-Hermitian Hamiltonians $H_1$ and $H_2$ are topologically equivalent, there exists a continuous deformation path connecting them without closing the gap. 
Conversely, if $H_1$ and $H_2$ are topologically distinct, any continuous deformation between them will inevitably result in gap closing. 
Rather than searching for such a continuous path~\cite{Scheurer_2020}, we define an initial continuous path and subsequently assess its robustness by introducing symmetry-preserving perturbations~\cite{Long_2023}. 
In our approach, the continuous deformation between two Hamiltonians $H_1$ and $H_2$ is realized using linear interpolation: $H_\alpha = (1-\alpha) H_1 + \alpha H_2$, where $a\in [0,1]$. 
This method preserves the representation basis and ensures the symmetry conditions are maintained throughout the deformation process. 
To identify the topological difference between $H_1$ and $H_2$, we detect the presence of a topologically protected crossing point at the Fermi level $E_f$~\cite{Long_2023} within the interpolated Hamiltonian $H_\alpha$ . 
Since the eigenenergies of $H_{1(2)}$ can be arbitrary, we utilize the flattened Hamiltonian $Q_{1(2)}$ of $H_{1(2)}$.  
This allows us to detect the topologically protected crossing point in $Q_\alpha= (1-\alpha) Q_1 + \alpha Q_2$ rather than in $H_\alpha$~\cite{Note1}. 

We define the similarity function based on the flattened Hamiltonian $Q$. 
Although the similarity function depends on the gap type, it can be expressed in a compact form~\cite{Note1}.
For a non-Hermitian Hamiltonian $H$,  we consider the eigen-equations $H |\psi_{n,\bm{k}}\rangle = E_n |\psi_{n,\bm{k}}\rangle$ and $H^\dagger |\varphi_{n,\bm{k}}\rangle = E_n^* |\varphi_{n,\bm{k}}\rangle$. 
For a line gap, the projection operator of $n$-th band can be defined as: $P(\bm{k}) = \sum_{n \in cocc} |\psi_{n, \bm{k}}\rangle \langle \varphi_{n,\bm{k}}| $, where $cocc$ denotes the complex occupied bands. 
The flattened Hamiltonian $Q$ is then given by $Q(\bm{k}) = 1-2P(\bm{k})$. 
Here, $cocc$ depends on the type of the line gap: (1) for a real line gap, $cocc = \{n|{\rm Re}[E_n] < {\rm Re}[E_f] \}$; and (2) for an imaginary line gap, $cocc = \{n| {\rm Im}[E_n] < {\rm Im}[E_f] \}$. 
We define $v_{\rm line} = \Pi_n {\rm Re}[\lambda_n]$, where $\{\lambda_n\}$ are the eigenvalues of $Q_i(\bm{k})+Q_j(\bm{k})$. 
For a point gap, we define $\widetilde{H} = \begin{pmatrix}
0 & H-E_f \\
H^\dagger - E_f^* & 0
\end{pmatrix}$, which maps $H$ to a Hermitian Hamiltonian $\widetilde{H}$ with the emergent chiral symmetry,  $\sigma_z \widetilde{H}(\bm{k}) \sigma_z = - \widetilde{H} (\bm{k})$~\cite{Note1}. 
The projection operator for $\widetilde{H}$ is $\widetilde{P}(\bm{k}) = \sum_{n \in occ} |\widetilde{\varphi}_{n, \bm{k}}\rangle \langle \widetilde{\varphi}_{n,\bm{k}}| $, where $\widetilde{H} |\widetilde{\varphi}_{n, \bm{k}}\rangle = \widetilde{E}_n |\widetilde{\varphi}_{n, \bm{k}}\rangle $, and $occ= \{n|\widetilde{E}_n<0\}$. 
The flattened Hamiltonian $\widetilde{Q}$ is given by $\widetilde{Q}(\bm{k}) = 1-2\widetilde{P}(\bm{k})$. 
We define $v_{\rm point} = \Pi_n \widetilde{\lambda}_n$, where $\{\widetilde{\lambda}_n\}$ are the eigenvalues of $\widetilde{Q}_i(\bm{k})+\widetilde{Q}_j(\bm{k})$. 
The similarity function $\mathcal{K}_{ij} $ between the Hamiltonian sample $H_i$ and $H_j$ is then defined as:
\begin{equation}
\mathcal{K}^{\rm point/line}_{ij} = \prod_{k\in {\rm BZ}}  \left(1-\exp\left[-\frac{|v_{\rm point/line} |^2}{\varepsilon^2} \right] \right).
\label{eq:similarity_function}
\end{equation}
where $\varepsilon \in \mathbb{R}$, the subscript ${\rm point/real/imag}$ denotes the gap type. 
Based on the Eq.~\ref{eq:similarity_function}, the corresponding distance function is $d_{ij} = 1 - \mathcal{K}_{ij}$. 
During the calculation, symmetry-preserving perturbations are introduced to test the robustness of the crossing point~\footnote{See more details in Supplementary Material about: (1) The similarity functions for the different gap types; (2) Theoretical analysis of non-Hermitian topological systems demonstrated in the main text and related details in machine learning; (3) The scheme for generating random non-Hermitian Hamiltonians according to symmetry conditions; (4) The classification results of non-Hermitian Hamiltonians in the different dimension. 
(5) The discussions about the boundary effect and GBZ. (6) The discussions about the relation of Eq.~\ref{eq:mapping_classification} based on Clifford algebra. }. 
In practice, we set $\varepsilon \rightarrow 0$, so that the similarity function becomes a binary function: $\mathcal{K}_{ij} = 1$ for topologically equivalent Hamiltonians, and $\mathcal{K}_{ij} = 0$ for topologically distinct ones~\cite{Long_2023}. 

\begin{figure}[h]
\centering
\includegraphics[width=\linewidth]{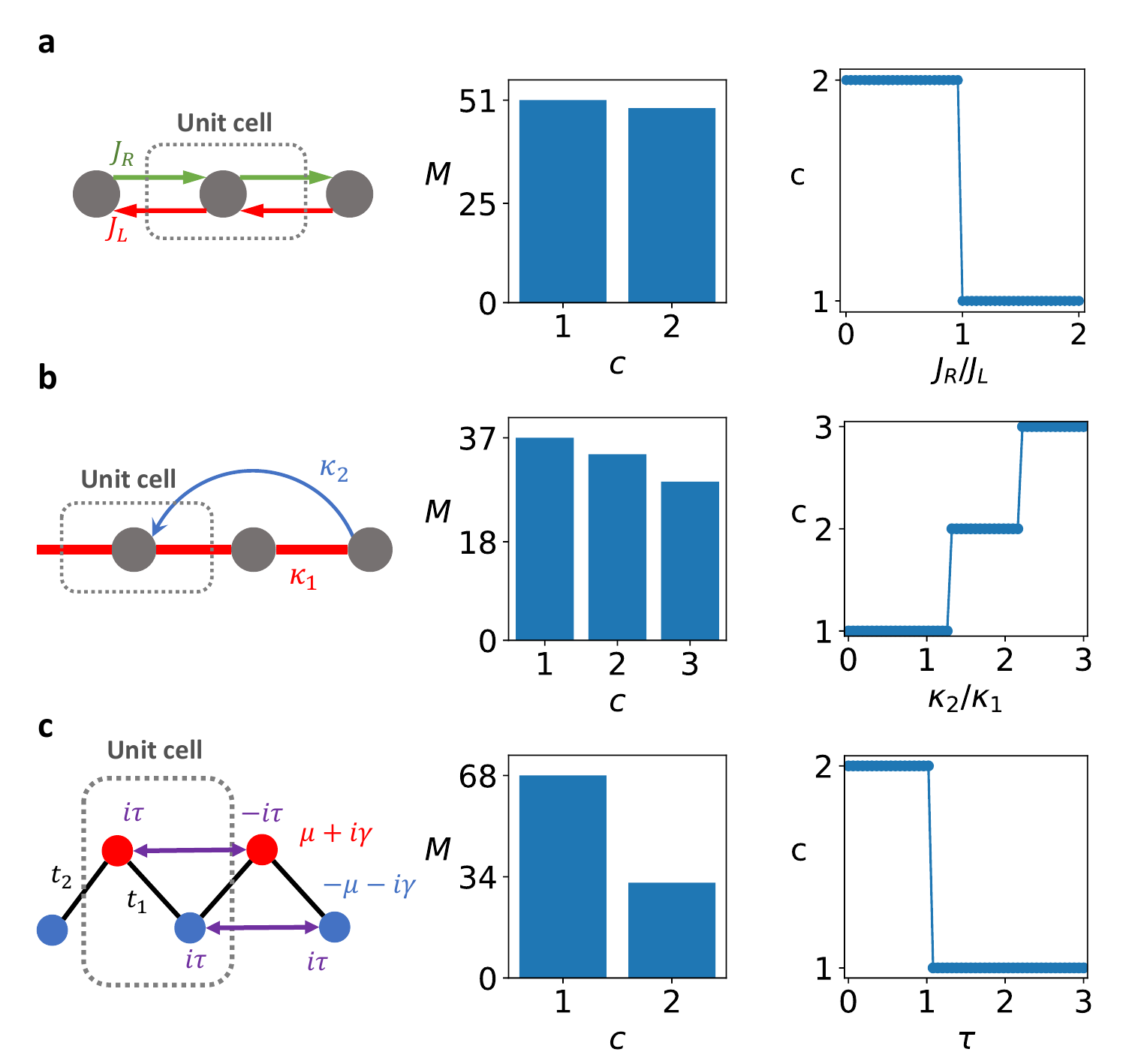}
\caption{Unsupervised learning of non-Hermitian point-gap topological phases. The left plots represent the system settings. The central plots represent the number of samples $M$ for the topologically distinct phases. $c$ denotes the custom label of phase. The right plots show the topological phase diagrams obtained by the similarity between the Hamiltonian with the given parameters and Hamiltonians in $\mathcal{G}$. The different colors and labels denote the topologically distinct phases, but not the topological invariants. (a) 1D Hatano-Nelson system. We set $J_L=1$, $J_R \in [0,2]$, $E_f=0$. (b) 1D non-Hermitian system with twisted winding in the complex-energy plane. We set $\kappa=1$, $\kappa_2 \in [0,3]$, $E_f=i$. (c) 1D non-Hermitian topological point-gap phase induced by onsite losses and gains. We set $t_1=\gamma=2$, $t_2=\mu=1$ and $\tau \in [0,3]$, $E_f=0$. Here, we generate 100 samples for each case. }
\label{fig:cases_point_gap}
\end{figure}

\begin{figure}[h]
\centering
\includegraphics[width=\linewidth]{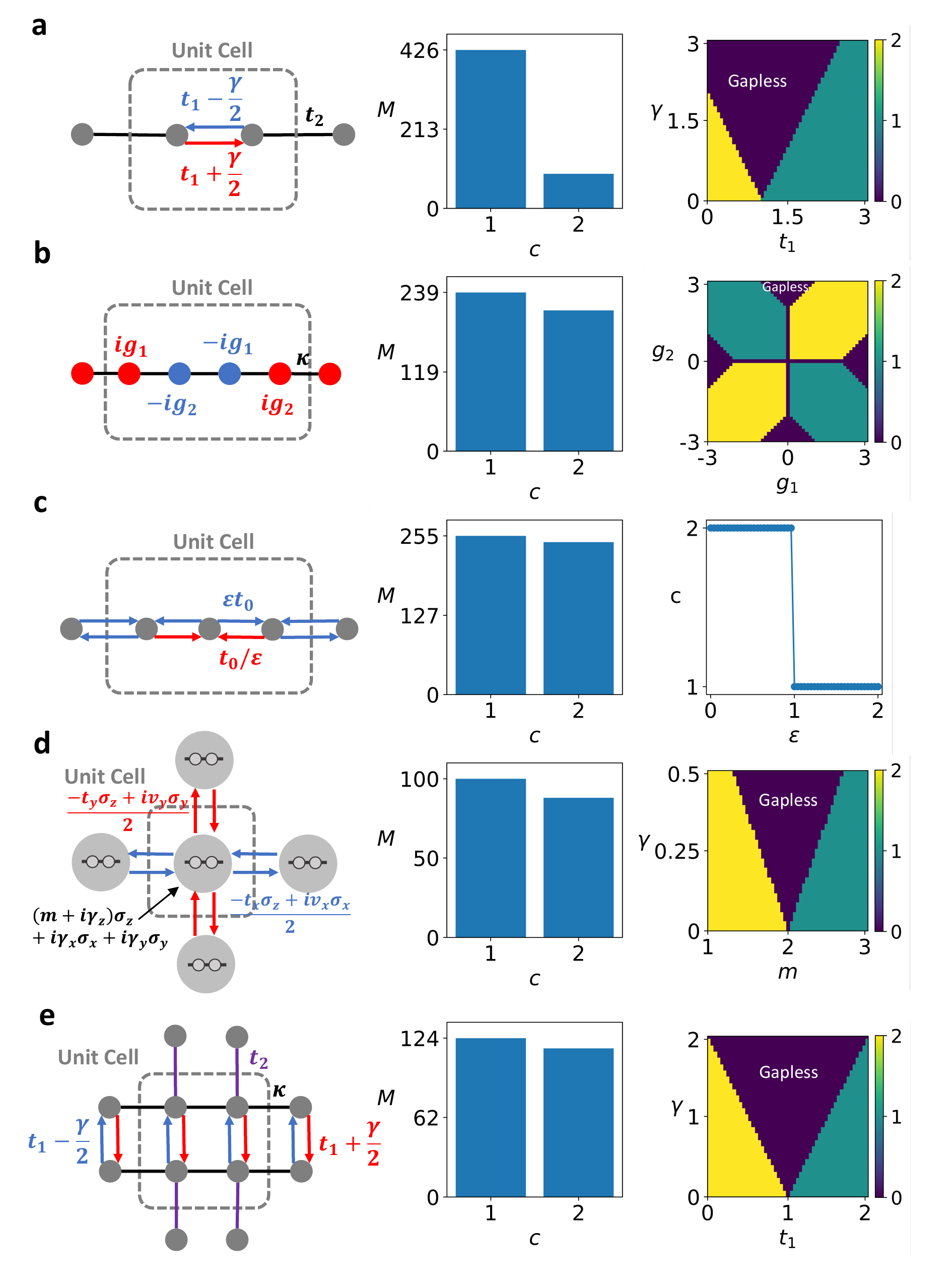}
\caption{Unsupervised learning of non-Hermitian real line-gap topological phases. The left plots represent the system settings. The central plots represent the number of samples $M$ for the topologically distinct phases. $c$ denotes the labels of phases. The right plots show the topological phase diagrams obtained by the similarity between the Hamiltonian with the given parameters and Hamiltonians in $\mathcal{G}$. The different colors and labels denote the topologically distinct phases but not the topological invariants. Note that $c=0$ denotes the gapless system.  (a) 1D non-Hermitian SSH system. We set $t_2=1$, $t_1\in [0,3]$, $\gamma\in [0,3]$ (b) 1D topological insulator phase solely induced by on-site gains and losses. We set $\kappa=1$, $g_1,g_2\in [-3,3]$. (c) 1D topological insulator phase solely induced by non-reciprocal couplings. We set $t_0=1$, $\varepsilon \in [0,2]$. (d) 2D non-Hermitian Chern insulator. We set $t_x=t_y=v_x=v_y=1$, $m\in [1,3]$, $\gamma \in [0,0.5]$. (e) 2D non-Hermitian topological M\"obius insulator. We set $\kappa=0.25$, $t_2=1$, $t_1 \in [0,2]$, $\gamma \in [0,2]$.
Here,  for obtaining the central plots, we generate 500 samples for each case and filter out the gapless systems. $E_f=0$ for all cases. }
\label{fig:cases_line_gap}
\end{figure}

We apply our previously proposed clustering algorithm in Ref.~\cite{Long_2023} to detect the number of phases and identify the phases in the Hamiltonian samples $\{ H_i \}$. 
Below, we provide a brief overview of the algorithm.
The algorithm operates on a set $\mathcal{G}$ and a list $\{M_c\}$, where $\mathcal{G} = \{H_{p_c}\}$ is a set of samples that are mutually different (i.e., $\mathcal{K}_{p_c p_{c'}}<1/2$, $\forall H_{p_c}, H_{p_{c'}}\in \mathcal{G}$) and $M_c$ denotes the number of samples that are topologically equivalent to $H_{p_c}$, $\{M_c|c=1,2,...N_c\}$. 
The algorithm proceeds in two steps: (1) The first sample $H_1$ is added into $\mathcal{G}$ since the initial $\mathcal{G} = \emptyset$. Then, $\mathcal{G}=\{H_1\}$, $p_1=1$, $M_1 = 1$ and $N_c=1$. 
(2) For each subsequently sample $H_j$, the algorithm compares it with the samples in $\mathcal{G}$. If $H_j$ is topologically equivalent to $H_{p_c}$, i.e., $H_{p_c}\in \mathcal{S}$ and $\mathcal{K}_{j,p_c}> \kappa_c$,  then $M_c := M_c + 1$. Otherwise, if none of the samples in $\mathcal{G}$ is topologically equivalent to $H_j$, $H_j$ is added into  $\mathcal{G}$, $M_{N_c+1} = 1$, $p_{N_c+1}=j$ and $N_c := N_c + 1$. 
After processing all the samples in $\{ H_i \}$, we can obtain: $N_c$ denotes the number of topologically distinct phases, and $\{M_c\}$ denotes the number of samples that have the same phase as $H_{p_c}$~\cite{Note1}. The index $c$ is used to label the topologically distinct phases. 

In the following, we demonstrate the validity of our algorithm. 
Firstly, we apply the algorithm to identify the non-Hermiticity-induced point-gap topological phases. Typical cases include the 1D Hatano-Nelson model in Fig.~\ref{fig:cases_point_gap}(a)~\cite{Hatano_1996, Bergholtz_2021}, a 1D non-Hermitian system with twisted loop in Fig.~\ref{fig:cases_point_gap}(b)~\cite{Zhang_2021}, and a loss-and-gain induced point-gap topological system in Fig.~\ref{fig:cases_point_gap}(c)~\cite{Yi_2020}. 
For each mode, we generate samples with random
parameters, calculate their similarities, and apply the clustering algorithm to determine the number of topologically distinct phases. 
We can see that the number of topologically distinct phases, denoted as $N_c$, is found to be: $N_c=2$ in Fig.~\ref{fig:cases_point_gap}(a), $N_c=3$ in Fig.~\ref{fig:cases_point_gap}(b), and $N_c = 2$ in Fig.~\ref{fig:cases_point_gap}(c). 
After labelling all the phases (i.e., assigning different values of $c$), we calculate their similarities with the samples and classify samples based on the label of the sample $H_{p_c}$ which has the maximum similarity. 
Consequently, we can obtain the topological phase diagrams in an unsupervised manner, as shown in Fig.~\ref{fig:cases_point_gap}. 
These results align well with theoretical predictions~\cite{Note1}. 

Secondly, non-Hermiticity can also induce new topological phases in systems with a line gap~\cite{Takata2018, Gao_2020, Luo2019, Gao2021, Xue_2020}. 
Below we take the real line gap as an example. 
We apply our algorithm to systems with the real line-gap topology, including the 1D non-Hermitian Su–Schrieffer–Heeger(SSH) system in Fig.~\ref{fig:cases_line_gap}(a)~\cite{Yao_2018}, the 1D topological system with on-site gain and loss in Fig.~\ref{fig:cases_line_gap}(b)~\cite{Takata2018}, the 1D topological system with non-reciprocal couplings in  Fig.~\ref{fig:cases_line_gap}(c)~\cite{Long_2022}, the 2D non-Hermitian Chern insulator in Fig.~\ref{fig:cases_line_gap}(d)~\cite{Yao2018}, and the 2D non-Hermitian topological M\"obius insulator in Fig.~\ref{fig:cases_line_gap}(e)~\cite{Note1}. 
After generating random parameter samples for each model, we calculate their similarities and perform the clustering algorithm to determine the number of topologically distinct phases. 
In the same manner as before, we can obtain the topological phase diagrams unsupervisedly, as shown in Fig.~\ref{fig:cases_line_gap}. 
These results are in good agreement with theoretical predictions~\cite{Note1}.  

\begin{figure*}[tp!]
\centering
\includegraphics[width=\linewidth]{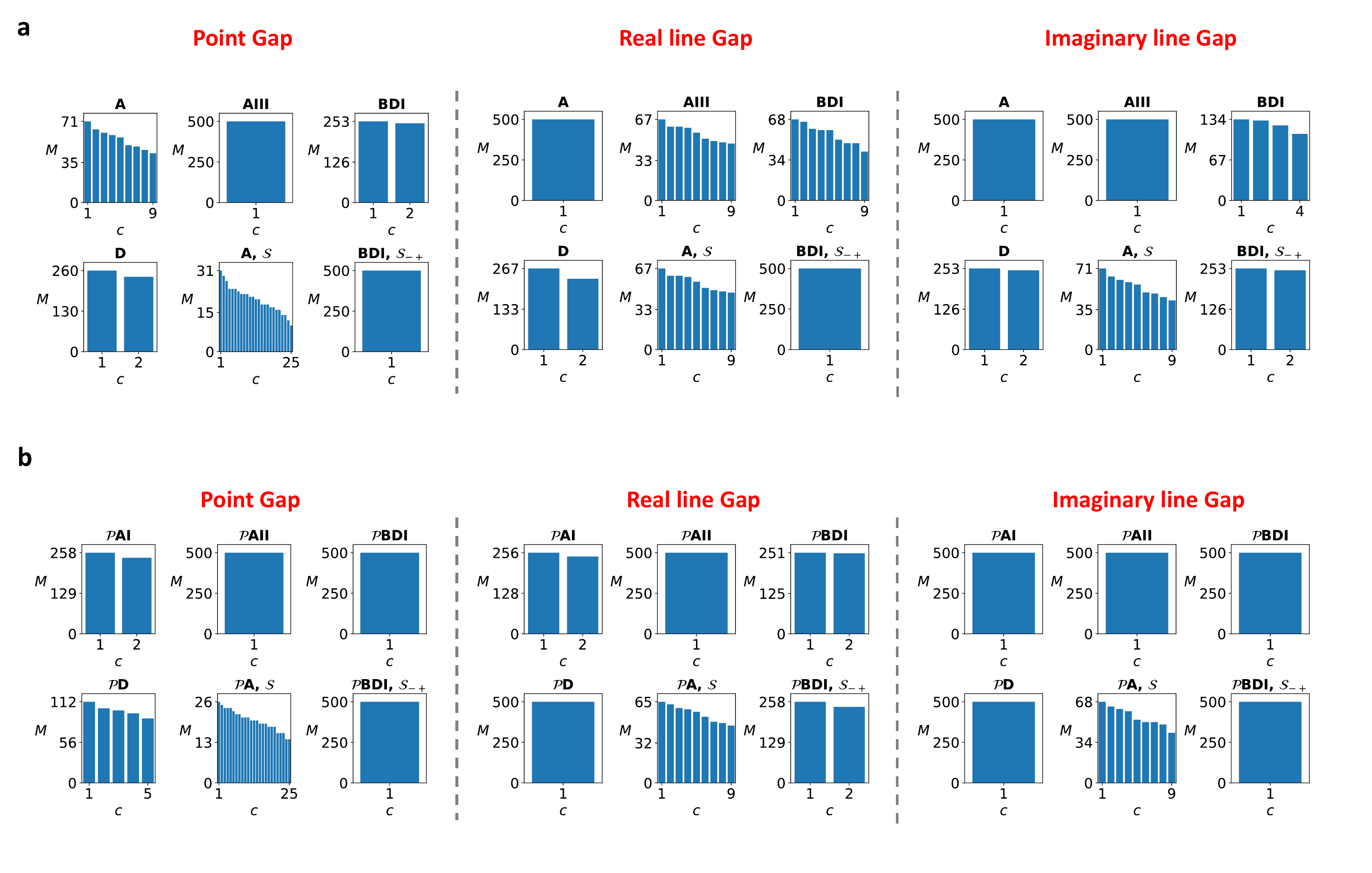}
\caption{Unsupervised learning of topological classifications of 1D non-Hermitian Hamiltonians in different symmetry classes.  
(a) Classification results for non-Hermitian symmetry classes. 
Here, we demonstrate 6 symmetry classes for each type of gaps.
(b) Classification results for non-Hermitian symmetry classes with considering the parity transformation. 
Here, we randomly generate 500 Hamiltonian samples for each symmetry class according to the 0D $n\times n$ Hamiltonians~\cite{Note1}. 
The number of phases $N_c$ can reflect the topological classification, e.g., $N_c=2$ corresponds to $\mathbb{Z}_2$ and $N_c =n+1$ corresponds to $\mathbb{Z}$. 
Here, we set $n=8$ and $E_f=0$.}
\label{fig:periodic_table}
\end{figure*}

Here, we apply our algorithm to achieve topological classification of non-Hermitian systems within the non-Hermitian symmetry classes. 
Non-Hermiticity enriches the symmetry classes into 38 distinct classes~\cite{Bernard_2002, Gong_2018, Kawabata_2019}, defined by combinations of not only time-reversal ($\mathcal{T}_{\pm}$), particle-hole ($\mathcal{C}_{\pm}$), and chiral ($\Gamma$) symmetries, but also sublattice ($\mathcal{S}$) and pseudo-Hermiticity ($\eta$) symmetries. 
By randomly generating Hamiltonian samples for each symmetry class, we employ our algorithm to determine the number of topologically distinct phases~\cite{Note1} .  
The resulting number of distinct phases from these randomly generated Hamiltonian samples under symmetries reveals the topological classification and reconstructs the topological periodic table~\cite{Long_2023}.  
As shown in Fig.~\ref{fig:periodic_table}(a), we demonstrate unsupervised classification of 1D non-Hermitian systems for three types of gaps acrosss selected symmetry classes. 

\begin{figure}[h]
\centering
\includegraphics[width=\linewidth]{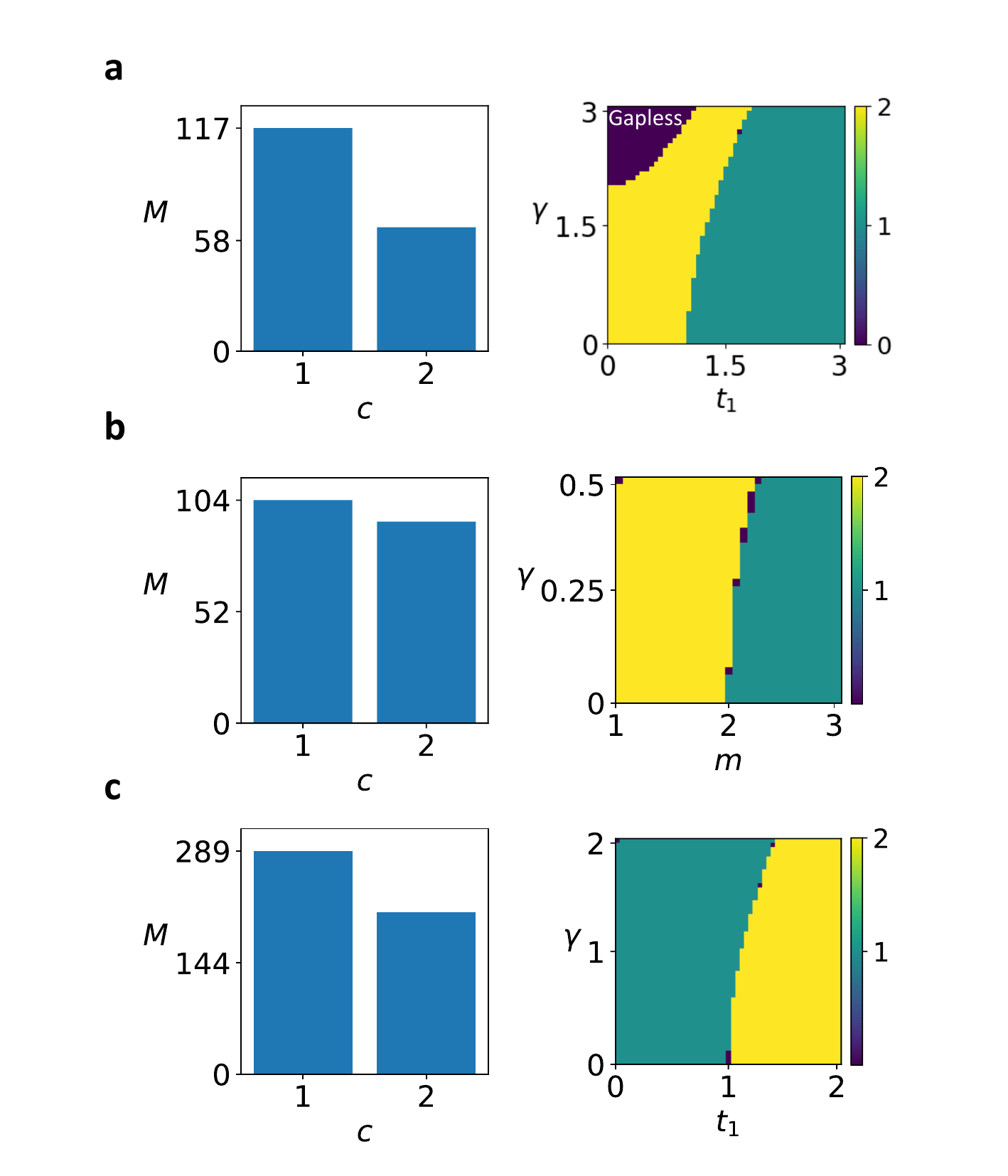}
\caption{Unsupervised learning of non-Hermitian real line-gap topological phases under GBZ. 
The left plots represent the number of samples $M$ for the topologically distinct phases. $c$ denotes the label of phases. The right plots show the topological phase diagrams obtained by the similarity between the Hamiltonian with the given parameters and Hamiltonians in $\mathcal{G}$. The different colors and labels denote the topologically distinct phases, but not the topological invariants. Note that $c=0$ denotes the gapless system. 
(a) 1D non-Hermitian SSH system. We set $t_2=1$, $t_1\in [0,3]$, $\gamma \in [0,3]$. 
(b) 2D non-Hermitian Chern insulator. We set $t_x=t_y=v_x=v_y=1$, $m\in [1,3]$, $\gamma \in [0, 0.5]$.  
(c) 2D non-Hermitian topological M\"obius insulator. We set $\kappa=0.25$, $t_2=1$, $t_1\in [0,2]$, $\gamma \in [0,2]$. 
Here, for obtaining the left plots, we randomly generate 500 samples for each case and filter out the gapless systems. $E_f=0$ for all cases. 
}
\label{fig:cases_OBC}
\end{figure}

From our calculations, we summarize the topological classification results for non-Hermitian systems under symmetries in Table.~\ref{tab:periodic_table}. 
Higher-dimensional non-Hermitian Hamiltonians are generated based on 0D $n$-banded Hamiltonians~\cite{Note1, Long_2023, Teo2010, Chiu_2016}. 
From Table.~\ref{tab:periodic_table}, we observe the following key features: (1) The classification exhibits an $8$-fold periodicity with respect to the dimension $d$, similar to the Bott periodicity in the topological periodic table for Hermitian systems. 
(2) The classification results for a given symmetry class can vary depending on the type of the gap considered.  
The topological classification can be deduced from the number of phases $N_c$~\cite{Long_2023,  Note1} as follows: (1) $N_c=1$ corresponds to a trivial group; (2) $N_c=n+1$ corresponds to $\mathbb{Z}$; (3) $N_c=n/2 + 1$ corresponds to $2\mathbb{Z}$; (4) $N_c=2$ corresponds to $\mathbb{Z}_2$; (5)  $N_c=(\frac{n}{2}+1)^2$ corresponds to $\mathbb{Z}\oplus\mathbb{Z}$; (6)  $N_c=(\frac{n}{4} + 1)^2$ corresponds to $2\mathbb{Z}\oplus 2\mathbb{Z}$; (7) $N_c=2\times 2$ corresponds to $\mathbb{Z}_2 \oplus \mathbb{Z}_2$. 
Clearly, our classification results are in good agreements with the theoretical predictions based on the homotopy groups of classifying space for abstract Hamiltonians~\cite{Gong_2018, Kawabata_2019, Wojcik_2020}, Clifford algebra~\cite{Zhou_2019} or topological field theory~\cite{Kawabata2021}. 
More details can be found in the Supplementary Material~\cite{Note1}. 

Parity-time ($\mathcal{PT}$) symmetry, defined as $\mathcal{PT}: \bm{r}\rightarrow -\bm{r}$, $t\rightarrow -t$, is a fundamental symmetry in non-Hermitian systems and plays a pivotal role in the development of non-Hermitian devices~\cite{Rueter2010, Mostafazadeh2002,ElGanainy2018,Oezdemir2019,Miri2019,Ding2022,Li2023a}. 
To investigate how parity transformation affects topological classification, we combine symmetry conditions with the parity transformation $\mathcal{P}$ ($\bm{r} \rightarrow -\bm{r}$)~\cite{Note1}. 
This transformation modifies $\mathcal{T}$ symmetry and $\mathcal{C}$ symmetry into $\mathcal{PT}$ symmetry and $\mathcal{PC}$ symmetry~\cite{Kawabata2019a}, respectively.
In Fig.~\ref{fig:periodic_table}(b), we present the topological classification results of Hamiltonian samples in symmetry classes after incorporating the parity transformation. 
For example, class $\mathcal{P}$AII corresponds to Hamiltonians with $\mathcal{PT}$ symmetry. 
The results, summarized in Table.~\ref{tab:periodic_table_parity}, reveal clear differences from Table.~\ref{tab:periodic_table}, indicating that $\mathcal{P}$ alters the topological classifications. 
Notably, there exists a systematic correspondence between classifications before and after before and after applying $\mathcal{P}$, expressed by the relation~\cite{Note1}: 
\begin{equation}
K^{\mathcal{P}}_{d} (s^{\mathcal{P}}) = K_{8-d} (s)
\label{eq:mapping_classification}
\end{equation}
where $K_{d}(s)$ represents the topological classification of $d$-dimensional non-Hermitian systems in symmetry classes $s$, and the superscript $\mathcal{P}$ denotes the classification after performing $\mathcal{P}$. 
For example, if $s$ corresonds to class AII, then $s^{\mathcal{P}}$ corresponds to $\mathcal{P}$AII. 
The mapping in Eq.~\ref{eq:mapping_classification} highlights that the introduction of $\mathcal{P}$ reverses the periodicity of the classifications (analogous to Bott periodicity) without introducing new topological phases. 
Crucially, this reversal is nontrivial: the parity transformation $\mathcal{P}$ aligns the topological classification of lower-dimensional systems with that of  higher-dimensional systems. 
For example, 1D Hamiltonians in class $\mathcal{P}$AII ($\mathcal{PT}$ symmetry) with a real line gap exhibit the same topological classification, $\mathbb{Z}_2$ ($N_c=2$), as 7D Hamiltonians in class AII. 

Finally, we explore the effects of open boundaries on non-Hermitian topological phases. Unlike Hermitian systems, where the conventional bulk-boundary correspondence (BBC) reliably connects bulk properties to boundary states, non-Hermitian systems often deviate from this principle due to unique features of non-Hermitian topology~\cite{Kunst2018, Yao_2018,Song_2019,Bergholtz_2021}. These deviations arise primarily from the nontrivial point-gap topology and manifest in several ways: (1) the non-Hermitian skin effect (NHSE)~\cite{Gong_2018, Okuma2020, Li_2020, Zhang_2021,Zhang_2022, Bergholtz_2021, Zhang_2021}, in which the extended Bloch states under periodic boundary conditions (PBC) become localized under open boundary conditions (OBC); (2) a mismatch between bulk and finite-size spectral gaps, such that a system lacking a line gap under PBC (e.g., being gapless) can exhibit a line gap under OBC~\cite{Lee_2016}; (3) shifts in the topological phase transition boundaries, where the phase diagram of the line-gap topology changes significantly with boundary conditions~\cite{Yao_2018, Yao2018}. 
In contrast, non-Hermitian  systems with trivial point-gap topology retain the conventional BBC~\cite{Long_2022}. 
These open-boundary effects can be described using the generalized Brillouin zone (GBZ) formalism, where the Bloch phase factor $e^{ik}$ is replaced by a complex number $\beta$~\cite{Yao_2018, Yao2018}.  
The shape and size of the GBZ depend sensitively on the system's parameters, providing a crucial framework for analyzing boundary effects in non-Hermitian systems.

Here, we demonstrate that our algorithm can operate under the GBZ to account for open-boundary effects~\cite{Note1}. 
We examine several non-Hermitian systems, including the 1D non-Hermitian SSH model in Fig.~\ref{fig:cases_OBC}(a)~\cite{Yao_2018}, the 2D non-Hermitian Chern insulator in Fig.~\ref{fig:cases_OBC}(b)~\cite{Yao2018}, and the 2D non-Hermitian topological M\"obius insulator in Fig.~\ref{fig:cases_OBC}(c)~\cite{Note1}. 
By generating random parameter samples for each model, we compute their similarities and apply the clustering algorithm to determine the number of topologically distinct phases. 
The resulting topological phase diagrams and the number of topologically distinct phases are shown in Fig.~\ref{fig:cases_OBC}. 
Notably, although the open-boundary effects can break the conventional BBC, they do not introduce new topological phases for real line-gap topology. 
As expected, the number of phases under OBC matches that under PBC in Fig.~\ref{fig:cases_line_gap}, excluding gapless phases. 
The classification results agree well with theoretical predictions~\cite{Note1}. 

For non-Hermitian systems exhibiting both point-gap and line-gap topology simultaneously, a key feature is the non-zero shift in the topological phase transition point or boundary of line-gap topological phases upon changing boundary conditions~\cite{Bergholtz_2021, Yao_2018,Yao2018,Song_2019,Lee2019}. 
However, not all symmetry conditions allow the coexistence of non-trivial point-gap and line-gap topology. 
Here, we take the line-gap topology in 1D systems as an example. 
According to Tables.~\ref{tab:periodic_table} and ~\ref{tab:periodic_table_parity}, we can conclude that the Hamiltonians in the following symmetry classes can have the non-zero shift of topological phase transition point/boundary after changing boundary conditions: 
AI, BDI, D, DIII$^\dagger$, AIII with $\mathcal{S}_+$, A with $\mathcal{S}$,  BDI with $\mathcal{S}_{++}$, DIII with $\mathcal{S}_{--}$, CII with $\mathcal{S}_{++}$, AI with $\mathcal{S}_{-}$, D with $\mathcal{S}_{+}$, C with $\mathcal{S}_{+}$, DIII with $\mathcal{S}_{++}$, CI with $\mathcal{S}_{++}$, AI with $\mathcal{S}_{+}$, BDI with $\mathcal{S}_{+-}$, D with $\mathcal{S}_{-}$, AII with $\mathcal{S}_{+}$, $\mathcal{P}$AI, $\mathcal{P}$CI, $\mathcal{P}$AI$^\dagger$, $\mathcal{P}$BDI$^\dagger$, $\mathcal{P}$BDI+$\mathcal{S}_{++}$, $\mathcal{P}$DIII with $\mathcal{S}_{--}$, $\mathcal{P}$CI with $\mathcal{S}_{--}$, $\mathcal{P}$AI with $\mathcal{S}_{-}$, $\mathcal{P}$D with $\mathcal{S}_{+}$, $\mathcal{P}$BDI with $\mathcal{S}_{+-}$, $\mathcal{P}$CII with $\mathcal{S}_{+-}$, $\mathcal{P}$C with $\mathcal{S}_{-}$, $\mathcal{P}$CI with $\mathcal{S}_{+-}$. 

To summarize, we propose an algorithm for the unsupervised topological classification of non-Hermitian topological systems under symmetries.  
This algorithm distinguishes topological differences among non-Hermitian Hamiltonians with symmetries, without relying on topological invariants, thereby avoiding the limitations associated with topological invariants. A topological periodic table for non-Hermitian systems across different symmetry classes is constructed in an unsupervised manner. 
Additionally, we incorporate unsupervised learning based on the GBZ to account for boundary effects. 
Our work paves the unsupervised way to identify the non-Hermitian topological phase, obtain the topological classification and guide new non-Hermitian topological devices~\cite{Harari2018, Bandres2018, Zeng2020, Amelio2020, Shao2019}. 
Furthermore, this approach can be extended to classify non-Hermitian Hamiltonians with other symmetries, such as dissipative symmetries in third quantization of open quantum systems~\cite{McDonald2023} and global symmetries in quadratic Lindbladians~\cite{Lieu2020}. 
Our work can also be extended to identify topological phases of interacting systems, if interacting systems can be described by an effective non-interacting Hamiltonians (i.e., by mean-field theories or quasi-particle representations~\cite{Kitaev2001, Wang2025}). 

\textbf{Code and data availability}. The source code for our implementation is available
in Ref.~\cite{long_github_2024_non_Hermitian}.

\begin{acknowledgments}
This research is supported by Singapore National Research Foundation Competitive Research Program under Grant no. NRF-CRP23-2019-0007, and Singapore Ministry of Education Academic Research Fund Tier 2 under Grant No. MOE-T2EP50123-0007. 
H.X. acknowledges the support from the National Natural Science Foundation of China under Grant No. 62401491, the Chinese University of Hong Kong under Grants No. 4937205, 4937206, 4053729 and 4411765, and the Research Grants Council of the Hong Kong Special Administrative Region, China under Grant No. 24304825. 
Y.L. gratefully acknowledges the support of the Eric and Wendy Schmidt AI in Science Postdoctoral Fellowship, a Schmidt Futures program. 
\end{acknowledgments}

\begin{table*}[h]
	\centering
	\caption{The number of topologically different phases $N_c$ for the $d$-dimensional non-Hermitian Hamiltonians in different symmetry classes. Non-Hermitian topological phases are classified according to the symmetry classes, the dimension $d$, and the definition of complex-energy point (P) or line (L) gaps. The subscript of L denotes the line gap for the real or imaginary part of the complex spectrum. $\mathcal{S}$ denotes the sublattice symmetry. The subscript of $\mathcal{S}_{\pm}$ denotes the commutation/anti-commutation relation to time-reversal $\mathcal{T}$ or particle-hole $\mathcal{C}$ symmetry. When $\mathcal{T}$ and $\mathcal{C}$ symmetry coexist, the first sign specifies the relation to $\mathcal{T}$ symmetry and the second sign to $\mathcal{C}$ symmetry.  Here, we generate the higher-dimensional Hamiltonians based on the 0D $n \times n$ random Hamiltonians~\cite{Note1}. \\}
	\label{tab:periodic_table}
     \begin{tabular}{cccccccccccc} \hline \hline
    ~Symmetry class~ & ~Gap~  & ~$d=0$~ & ~$d=1$~ & ~$d=2$~ & ~$d=3$~ & ~$d=4$~ & ~$d=5$~ & ~$d=6$~ & ~$d=7$~ & ~$d=8$~ & ~$d=9$~\\ \hline
    \multirow{3}{*}{A}
    & P & \Cb \\
    & \Lr & \Ca \\ 
    & \Li & \Ca \\ \hline
    \multirow{3}{*}{AIII}
    & P & \Ca \\    
    & \Lr & \Cb \\ 
    & \Li & \Caa \\ \hline 
     \multirow{3}{*}{AI}
    & P & \Rb \\    
    & \Lr & \Ra \\ 
    & \Li & \Rc \\ \hline
    \multirow{3}{*}{BDI}
    & P & \Rc \\    
    & \Lr & \Rb \\ 
    & \Li & \Rcc \\ \hline
    \multirow{3}{*}{D}
    & P & \Rd \\    
    & \Lr & \Rc \\ 
    & \Li & \Rc \\ \hline
    \multirow{3}{*}{DIII}
    & P & \Ree \\    
    & \Lr & \Rd \\ 
    & \Li & \Ca \\ \hline
    \multirow{3}{*}{AII}
    & P & \Rf \\    
    & \Lr & \Ree \\ 
    & \Li & \Rg \\ \hline
    \multirow{3}{*}{CII}
    & P & \Rg \\    
    & \Lr & \Rf \\ 
    & \Li & \Rgg \\ \hline
    \multirow{3}{*}{C}
    & P & \Rh \\    
    & \Lr & \Rg \\ 
    & \Li & \Rg \\ \hline
    \multirow{3}{*}{CI}
    & P & \Ra \\    
    & \Lr & \Rh \\ 
    & \Li & \Ca \\ \hline
    \multirow{3}{*}{AI$^\dagger$}
    & P & \Rh \\    
    & \Lr & \Ra \\ 
    & \Li & \Ra \\ \hline
    \multirow{3}{*}{BDI$^\dagger$}
    & P & \Ra \\    
    & \Lr & \Rb \\ 
    & \Li & \Raa \\ \hline
     \multirow{3}{*}{DIII$^\dagger$}
    & P & \Rc \\    
    & \Lr & \Rd \\ 
    & \Li & \Ca \\ \hline
    \multirow{3}{*}{AII$^\dagger$}
    & P & \Rd \\    
    & \Lr & \Ree \\ 
    & \Li & \Ree \\ \hline
    \multirow{3}{*}{CII$^\dagger$}
    & P & \Ree \\    
    & \Lr & \Rf \\ 
    & \Li & \Reee \\ \hline
    \multirow{3}{*}{CI$^\dagger$}
    & P & \Rg \\    
    & \Lr & \Rh \\ 
    & \Li & \Ca \\ \hline
    \hline
     \multicolumn{12}{c}{continued on next page}
  \end{tabular}
\end{table*}

\begin{table*}[h]
	\centering
     \begin{tabular}{cccccccccccc}
    \multicolumn{12}{c}{Table \ref{tab:periodic_table} --- continued} \\ 
     \hline \hline
    ~Symmetry class~ & ~Gap~  & ~$d=0$~ & ~$d=1$~ & ~$d=2$~ & ~$d=3$~ & ~$d=4$~ & ~$d=5$~ & ~$d=6$~ & ~$d=7$~ & ~$d=8$~ & ~$d=9$~\\ \hline
   \multirow{3}{*}{AIII, $\mathcal{S}_{+}$}
    & P & \Cb \\    
    & \Lr & \Cbb \\ 
    & \Li & \Cbb \\ \hline
    \multirow{3}{*}{A, $\mathcal{S}$}
    & P & \Cbb \\    
    & \Lr & \Cb \\ 
    & \Li & \Cb \\ \hline
    \multirow{3}{*}{AIII, $\mathcal{S}_{-}$}
    & P & \Caa \\    
    & \Lr & \Ca \\ 
    & \Li & \Ca \\ \hline
    \multirow{3}{*}{BDI, $\mathcal{S}_{++}$}
    & P & \Rb \\    
    & \Lr & \Rbb \\ 
    & \Li & \Rbb \\ \hline
     \multirow{3}{*}{DIII, $\mathcal{S}_{--}$}
    & P & \Rd \\    
    & \Lr & \Rdd \\ 
    & \Li & \Cb \\ \hline
    \multirow{3}{*}{CII, $\mathcal{S}_{++}$}
    & P & \Rf \\    
    & \Lr & \Rff \\ 
    & \Li & \Rff \\ \hline
    \multirow{3}{*}{CI, $\mathcal{S}_{--}$}
    & P & \Rh \\    
    & \Lr & \Rhh \\ 
    & \Li & \Cb \\ \hline
    \multirow{3}{*}{AI, $\mathcal{S}_{-}$}
    & P & \Cb \\    
    & \Lr & \Rh \\ 
    & \Li & \Rd \\ \hline
     \multirow{3}{*}{BDI, $\mathcal{S}_{-+}$}
    & P & \Ca \\    
    & \Lr & \Ra \\ 
    & \Li & \Rc \\ \hline
    \multirow{3}{*}{D, $\mathcal{S}_{+}$}
    & P & \Cb \\    
    & \Lr & \Rb \\ 
    & \Li & \Rb \\ \hline
    \multirow{3}{*}{CII, $\mathcal{S}_{-+}$}
    & P & \Ca \\    
    & \Lr & \Ree \\ 
    & \Li & \Rg \\ \hline
    \multirow{3}{*}{C, $\mathcal{S}_{+}$}
    & P & \Cb \\    
    & \Lr & \Rf \\ 
    & \Li & \Rf \\ \hline
    \multirow{3}{*}{DIII, $\mathcal{S}_{++}$}
    & P & \Rf \\    
    & \Lr & \Cb \\ 
    & \Li & \Cb \\ \hline
     \multirow{3}{*}{CI, $\mathcal{S}_{++}$}
    & P & \Rb \\    
    & \Lr & \Cb \\ 
    & \Li & \Cb \\ \hline
     \multirow{3}{*}{AI, $\mathcal{S}_{+}$}
    & P & \Rb \\    
    & \Lr & \Rb \\ 
    & \Li & \Rb \\ \hline
     \multirow{3}{*}{BDI, $\mathcal{S}_{+-}$}
    & P & \Rcc \\    
    & \Lr & \Rc \\ 
    & \Li & \Rc \\ \hline
    \multirow{3}{*}{D, $\mathcal{S}_{-}$}
    & P & \Rdd \\    
    & \Lr & \Rd \\ 
    & \Li & \Rd \\ \hline
     \hline
     \multicolumn{12}{c}{continued on next page}
  \end{tabular}
\end{table*}

\begin{table*}[h]
	\centering
     \begin{tabular}{cccccccccccc}
    \multicolumn{12}{c}{Table \ref{tab:periodic_table} --- continued} \\ 
     \hline \hline
    ~Symmetry class~ & ~Gap~  & ~$d=0$~ & ~$d=1$~ & ~$d=2$~ & ~$d=3$~ & ~$d=4$~ & ~$d=5$~ & ~$d=6$~ & ~$d=7$~ & ~$d=8$~ & ~$d=9$~\\ \hline
    \multirow{3}{*}{DIII, $\mathcal{S}_{+-}$}
    & P & \Reee \\    
    & \Lr & \Ree \\ 
    & \Li & \Ree \\ \hline
     \multirow{3}{*}{AII, $\mathcal{S}_{+}$}
    & P & \Rff \\    
    & \Lr & \Rf \\ 
    & \Li & \Rf \\ \hline
    \multirow{3}{*}{CII, $\mathcal{S}_{+-}$}
    & P & \Rgg \\    
    & \Lr & \Rg \\ 
    & \Li & \Rg \\ \hline
    \multirow{3}{*}{C, $\mathcal{S}_{-}$}
    & P & \Rhh \\    
    & \Lr & \Rh \\ 
    & \Li & \Rh \\ \hline
    \multirow{3}{*}{CI, $\mathcal{S}_{+-}$}
    & P & \Raa \\    
    & \Lr & \Ra \\ 
    & \Li & \Ra \\ \hline
     \hline
  \end{tabular}
\end{table*}

\begin{table*}[h]
	\centering
	\caption{The number of topologically different phases $N_c$ for the $d$-dimensional non-Hermitian Hamiltonians in different parity-equipped symmetry classes. Non-Hermitian topological phases are classified according to the symmetry classes, the dimension $d$, and the definition of complex-energy point (P) or line (L) gaps. The subscript of L denotes the line gap for the real or imaginary part of the complex spectrum. $\mathcal{S}$ denotes the sublattice symmetry. The subscript of $\mathcal{S}_{\pm}$ denotes the commutation/anti-commutation relation to time-reversal $\mathcal{T}$ or particle-hole $\mathcal{C}$ symmetry. When $\mathcal{T}$ and $\mathcal{C}$ symmetry coexist, the first sign specifies the relation to $\mathcal{T}$ symmetry and the second sign to $\mathcal{C}$ symmetry. Note that the dimension $d$ starts from $d=1$, because 0D Hamiltonians don't have the parity transformation.  Class A and AIII are identical to the classes in Table.~\ref{tab:periodic_table}. Here, we generate the higher-dimensional Hamiltonians based on the 0D $n \times n$ random Hamiltonians~\cite{Note1}.\\}
	\label{tab:periodic_table_parity}
     \begin{tabular}{cccccccccccc} \hline \hline
    ~Symmetry class~ & ~Gap~  & ~$d=1$~ & ~$d=2$~ & ~$d=3$~ & ~$d=4$~ & ~$d=5$~ & ~$d=6$~ & ~$d=7$~ & ~$d=8$~ & ~$d=9$~ & ~$d=10$~\\ \hline
     \multirow{3}{*}{$\mathcal{P}$AI}
    & P & \PRc \\    
    & \Lr & \PRb \\ 
    & \Li & \PRd \\ \hline
    \multirow{3}{*}{$\mathcal{P}$BDI}
    & P & \PRd \\    
    & \Lr & \PRc \\ 
    & \Li & \PRdd \\ \hline
    \multirow{3}{*}{$\mathcal{P}$D}
    & P & \PRee \\    
    & \Lr & \PRd \\ 
    & \Li & \PRd \\ \hline
    \multirow{3}{*}{$\mathcal{P}$DIII}
    & P & \PRf \\    
    & \Lr & \PRee \\ 
    & \Li & \PCb \\ \hline
    \multirow{3}{*}{$\mathcal{P}$AII}
    & P & \PRg \\    
    & \Lr & \PRf \\ 
    & \Li & \PRh \\ \hline
    \multirow{3}{*}{$\mathcal{P}$CII}
    & P & \PRh \\    
    & \Lr & \PRg \\ 
    & \Li & \PRhh \\ \hline
    \multirow{3}{*}{$\mathcal{P}$C}
    & P & \PRa \\    
    & \Lr & \PRh \\ 
    & \Li & \PRh \\ \hline
    \multirow{3}{*}{$\mathcal{P}$CI}
    & P & \PRb \\    
    & \Lr & \PRa \\ 
    & \Li & \PCb \\ \hline
    \multirow{3}{*}{$\mathcal{P}$AI$^\dagger$}
    & P & \PRa \\    
    & \Lr & \PRb \\ 
    & \Li & \PRb \\ \hline
    \multirow{3}{*}{$\mathcal{P}$BDI$^\dagger$}
    & P & \PRb \\    
    & \Lr & \PRc \\ 
    & \Li & \PRbb \\ \hline
     \multirow{3}{*}{$\mathcal{P}$DIII$^\dagger$}
    & P & \PRd \\    
    & \Lr & \PRee \\ 
    & \Li & \PCb \\ \hline
    \multirow{3}{*}{$\mathcal{P}$AII$^\dagger$}
    & P & \PRee \\    
    & \Lr & \PRf \\ 
    & \Li & \PRf \\ \hline
    \multirow{3}{*}{$\mathcal{P}$CII$^\dagger$}
    & P & \PRf \\    
    & \Lr & \PRg \\ 
    & \Li & \PRff \\ \hline
    \multirow{3}{*}{$\mathcal{P}$CI$^\dagger$}
    & P & \PRh \\    
    & \Lr & \PRa \\ 
    & \Li & \PCb \\ \hline
    \hline
     \multicolumn{12}{c}{continued on next page}
  \end{tabular}
\end{table*}

\begin{table*}[h]
	\centering
     \begin{tabular}{cccccccccccc}
    \multicolumn{12}{c}{Table \ref{tab:periodic_table_parity} --- continued} \\ 
     \hline \hline
    ~Symmetry class~ & ~Gap~  & ~$d=1$~ & ~$d=2$~ & ~$d=3$~ & ~$d=4$~ & ~$d=5$~ & ~$d=6$~ & ~$d=7$~ & ~$d=8$~ & ~$d=9$ & ~$d=10$~\\ \hline
    \multirow{3}{*}{$\mathcal{P}$BDI, $\mathcal{S}_{++}$}
    & P & \PRc \\    
    & \Lr & \PRcc \\ 
    & \Li & \PRcc \\ \hline
     \multirow{3}{*}{$\mathcal{P}$DIII, $\mathcal{S}_{--}$}
    & P & \PRee \\    
    & \Lr & \PReee \\ 
    & \Li & \Ca \\ \hline
    \multirow{3}{*}{$\mathcal{P}$CII, $\mathcal{S}_{++}$}
    & P & \PRg \\    
    & \Lr & \PRgg \\ 
    & \Li & \PRgg \\ \hline
    \multirow{3}{*}{$\mathcal{P}$CI, $\mathcal{S}_{--}$}
    & P & \PRa \\    
    & \Lr & \PRaa \\ 
    & \Li & \PCa \\ \hline
    \multirow{3}{*}{$\mathcal{P}$AI, $\mathcal{S}_{-}$}
    & P & \PCa \\    
    & \Lr & \PRa \\ 
    & \Li & \PRee \\ \hline
     \multirow{3}{*}{$\mathcal{P}$BDI, $\mathcal{S}_{-+}$}
    & P & \PCb \\    
    & \Lr & \PRb \\ 
    & \Li & \PRd \\ \hline
    \multirow{3}{*}{$\mathcal{P}$D, $\mathcal{S}_{+}$}
    & P & \PCa \\    
    & \Lr & \PRc \\ 
    & \Li & \PRc \\ \hline
    \multirow{3}{*}{$\mathcal{P}$CII, $\mathcal{S}_{-+}$}
    & P & \PCb \\    
    & \Lr & \PRf \\ 
    & \Li & \PRh \\ \hline
    \multirow{3}{*}{$\mathcal{P}$C, $\mathcal{S}_{+}$}
    & P & \PCa \\    
    & \Lr & \PRg \\ 
    & \Li & \PRg \\ \hline
    \multirow{3}{*}{$\mathcal{P}$DIII, $\mathcal{S}_{++}$}
    & P & \PRg \\    
    & \Lr & \PCa \\ 
    & \Li & \PCa \\ \hline
     \multirow{3}{*}{$\mathcal{P}$CI, $\mathcal{S}_{++}$}
    & P & \PRc \\    
    & \Lr & \PCa \\ 
    & \Li & \PCa \\ \hline
     \multirow{3}{*}{$\mathcal{P}$AI, $\mathcal{S}_{+}$}
    & P & \PRcc \\    
    & \Lr & \PRc \\ 
    & \Li & \PRc \\ \hline
     \multirow{3}{*}{$\mathcal{P}$BDI, $\mathcal{S}_{+-}$}
    & P & \PRdd \\    
    & \Lr & \PRd \\ 
    & \Li & \PRd \\ \hline
    \multirow{3}{*}{$\mathcal{P}$D, $\mathcal{S}_{-}$}
    & P & \PReee \\    
    & \Lr & \PRee \\ 
    & \Li & \PRee \\ \hline
     \hline
      \multicolumn{12}{c}{continued on next page}
  \end{tabular}
\end{table*}

\begin{table*}[h]
	\centering
     \begin{tabular}{cccccccccccc}
    \multicolumn{12}{c}{Table \ref{tab:periodic_table_parity} --- continued} \\ 
     \hline \hline
    ~Symmetry class~ & ~Gap~  & ~$d=0$~ & ~$d=1$~ & ~$d=2$~ & ~$d=3$~ & ~$d=4$~ & ~$d=5$~ & ~$d=6$~ & ~$d=7$~ & ~$d=8$~ & ~$d=9$~\\ \hline
    \multirow{3}{*}{$\mathcal{P}$DIII, $\mathcal{S}_{+-}$}
    & P & \PRff \\    
    & \Lr & \PRf \\ 
    & \Li & \PRf \\ \hline
     \multirow{3}{*}{$\mathcal{P}$AII, $\mathcal{S}_{+}$}
    & P & \PRgg \\    
    & \Lr & \PRg \\ 
    & \Li & \PRg \\ \hline
    \multirow{3}{*}{$\mathcal{P}$CII, $\mathcal{S}_{+-}$}
    & P & \PRhh \\    
    & \Lr & \PRh \\ 
    & \Li & \PRh \\ \hline
    \multirow{3}{*}{$\mathcal{P}$C, $\mathcal{S}_{-}$}
    & P & \PRaa \\    
    & \Lr & \PRa \\ 
    & \Li & \PRa \\ \hline
    \multirow{3}{*}{$\mathcal{P}$CI, $\mathcal{S}_{+-}$}
    & P & \PRbb \\    
    & \Lr & \PRb \\ 
    & \Li & \PRb \\ \hline
     \hline
  \end{tabular}
\end{table*}

\bibliography{references}

@Article{Yu_2021,
  author    = {Li-Wei Yu and Dong-Ling Deng},
  journal   = {Phys. Rev. Lett.},
  title     = {Unsupervised Learning of Non-Hermitian Topological Phases},
  year      = {2021},
  month     = {jun},
  number    = {24},
  pages     = {240402},
  volume    = {126},
  doi       = {10.1103/physrevlett.126.240402},
  publisher = {American Physical Society ({APS})},
}

@Article{Long_2023,
  author    = {Yang Long and Baile Zhang},
  journal   = {Phys. Rev. Lett.},
  title     = {Unsupervised Data-Driven Classification of Topological Gapped Systems with Symmetries},
  year      = {2023},
  month     = {jan},
  number    = {3},
  pages     = {036601},
  volume    = {130},
  doi       = {10.1103/physrevlett.130.036601},
  publisher = {American Physical Society ({APS})},
}

@Article{Scheurer_2020,
  author    = {Mathias S. Scheurer and Robert-Jan Slager},
  journal   = {Phys. Rev. Lett.},
  title     = {Unsupervised Machine Learning and Band Topology},
  year      = {2020},
  month     = {jun},
  number    = {22},
  pages     = {226401},
  volume    = {124},
  doi       = {10.1103/physrevlett.124.226401},
  publisher = {American Physical Society ({APS})},
}

@Article{Yi_2020,
  author    = {Yifei Yi and Zhesen Yang},
  journal   = {Phys. Rev. Lett.},
  title     = {Non-Hermitian Skin Modes Induced by On-Site Dissipations and Chiral Tunneling Effect},
  year      = {2020},
  month     = {oct},
  number    = {18},
  pages     = {186802},
  volume    = {125},
  doi       = {10.1103/physrevlett.125.186802},
  publisher = {American Physical Society ({APS})},
}

@Article{Yao_2018,
  author    = {Shunyu Yao and Zhong Wang},
  journal   = {Phys. Rev. Lett.},
  title     = {Edge States and Topological Invariants of Non-Hermitian Systems},
  year      = {2018},
  month     = {aug},
  number    = {8},
  pages     = {086803},
  volume    = {121},
  doi       = {10.1103/physrevlett.121.086803},
  publisher = {American Physical Society ({APS})},
}

@Article{Lee_2016,
  author    = {Tony E. Lee},
  journal   = {Phys. Rev. Lett.},
  title     = {Anomalous Edge State in a Non-Hermitian Lattice},
  year      = {2016},
  month     = {apr},
  number    = {13},
  pages     = {133903},
  volume    = {116},
  doi       = {10.1103/physrevlett.116.133903},
  publisher = {American Physical Society ({APS})},
}

@Article{Zhang_2021,
  author    = {Li Zhang and Yihao Yang and Yong Ge and Yi-Jun Guan and Qiaolu Chen and Qinghui Yan and Fujia Chen and Rui Xi and Yuanzhen Li and Ding Jia and Shou-Qi Yuan and Hong-Xiang Sun and Hongsheng Chen and Baile Zhang},
  journal   = {Nat. Commun.},
  title     = {Acoustic non-Hermitian skin effect from twisted winding topology},
  year      = {2021},
  month     = {nov},
  number    = {1},
  pages     = {6297},
  volume    = {12},
  doi       = {10.1038/s41467-021-26619-8},
  publisher = {Springer Science and Business Media {LLC}},
}

@Article{Song_2019,
  author    = {Fei Song and Shunyu Yao and Zhong Wang},
  journal   = {Phys. Rev. Lett.},
  title     = {Non-Hermitian Topological Invariants in Real Space},
  year      = {2019},
  month     = {dec},
  number    = {24},
  pages     = {246801},
  volume    = {123},
  doi       = {10.1103/physrevlett.123.246801},
  publisher = {American Physical Society ({APS})},
}

@Article{Gong_2018,
  author    = {Zongping Gong and Yuto Ashida and Kohei Kawabata and Kazuaki Takasan and Sho Higashikawa and Masahito Ueda},
  journal   = {Phys. Rev. X},
  title     = {Topological Phases of Non-Hermitian Systems},
  year      = {2018},
  month     = {sep},
  number    = {3},
  pages     = {031079},
  volume    = {8},
  doi       = {10.1103/physrevx.8.031079},
  publisher = {American Physical Society ({APS})},
}

@Article{Kawabata_2019,
  author    = {Kohei Kawabata and Ken Shiozaki and Masahito Ueda and Masatoshi Sato},
  journal   = {Phys. Rev. X},
  title     = {Symmetry and Topology in Non-Hermitian Physics},
  year      = {2019},
  month     = {oct},
  number    = {4},
  pages     = {041015},
  volume    = {9},
  doi       = {10.1103/physrevx.9.041015},
  publisher = {American Physical Society ({APS})},
}

@Article{Bergholtz_2021,
  author    = {Emil J. Bergholtz and Jan Carl Budich and Flore K. Kunst},
  journal   = {Rev. Mod. Phys.},
  title     = {Exceptional topology of non-Hermitian systems},
  year      = {2021},
  month     = {feb},
  number    = {1},
  pages     = {015005},
  volume    = {93},
  doi       = {10.1103/revmodphys.93.015005},
  publisher = {American Physical Society ({APS})},
}

@Article{Hatano_1996,
  author    = {Naomichi Hatano and David R. Nelson},
  journal   = {Phys. Rev. Lett.},
  title     = {Localization Transitions in Non-Hermitian Quantum Mechanics},
  year      = {1996},
  month     = {jul},
  number    = {3},
  pages     = {570--573},
  volume    = {77},
  doi       = {10.1103/physrevlett.77.570},
  publisher = {American Physical Society ({APS})},
}

@Article{Li_2020,
  author    = {Linhu Li and Ching Hua Lee and Sen Mu and Jiangbin Gong},
  journal   = {Nat. Commun.},
  title     = {Critical non-Hermitian skin effect},
  year      = {2020},
  month     = {oct},
  number    = {1},
  pages     = {5491},
  volume    = {11},
  doi       = {10.1038/s41467-020-18917-4},
  publisher = {Springer Science and Business Media {LLC}},
}

@Article{Bernard_2002,
  author    = {Denis Bernard and Andr{\'{e}} LeClair},
  journal   = {Statistical Field Theories},
  title     = {A Classification of Non-Hermitian Random Matrices},
  year      = {2002},
  pages     = {207--214},
  booktitle = {Statistical Field Theories},
  doi       = {10.1007/978-94-010-0514-2_19},
  publisher = {Springer Netherlands},
}

@Article{Zhou_2019,
  author    = {Hengyun Zhou and Jong Yeon Lee},
  journal   = {Phys. Rev. B},
  title     = {Periodic table for topological bands with non-Hermitian symmetries},
  year      = {2019},
  month     = {jun},
  number    = {23},
  pages     = {235112},
  volume    = {99},
  doi       = {10.1103/physrevb.99.235112},
  publisher = {American Physical Society ({APS})},
}

@Article{Chiu_2016,
  author    = {Ching-Kai Chiu and Jeffrey C.{\hspace{0.167em}}Y. Teo and Andreas P. Schnyder and Shinsei Ryu},
  journal   = {Rev. Mod. Phys.},
  title     = {Classification of topological quantum matter with symmetries},
  year      = {2016},
  month     = {aug},
  number    = {3},
  pages     = {035005},
  volume    = {88},
  doi       = {10.1103/revmodphys.88.035005},
  publisher = {American Physical Society ({APS})},
}

@Article{Long_2022,
  author    = {Yang Long and Haoran Xue and Baile Zhang},
  journal   = {Phys. Rev. B},
  title     = {Non-Hermitian topological systems with eigenvalues that are always real},
  year      = {2022},
  month     = {mar},
  number    = {10},
  pages     = {l100102},
  volume    = {105},
  doi       = {10.1103/physrevb.105.l100102},
  publisher = {American Physical Society ({APS})},
}

@Article{Zhang_2022,
  author    = {Kai Zhang and Zhesen Yang and Chen Fang},
  journal   = {Nat. Commun.},
  title     = {Universal non-Hermitian skin effect in two and higher dimensions},
  year      = {2022},
  month     = {may},
  number    = {1},
  pages     = {2496},
  volume    = {13},
  doi       = {10.1038/s41467-022-30161-6},
  publisher = {Springer Science and Business Media {LLC}},
}

@Article{Lee2019,
  author    = {Ching Hua Lee and Linhu Li and Jiangbin Gong},
  journal   = {Phys. Rev. Lett.},
  title     = {Hybrid Higher-Order Skin-Topological Modes in Nonreciprocal Systems},
  year      = {2019},
  month     = {jul},
  number    = {1},
  pages     = {016805},
  volume    = {123},
  doi       = {10.1103/physrevlett.123.016805},
  publisher = {American Physical Society ({APS})},
}

@Article{Yao2018,
  author    = {Shunyu Yao and Fei Song and Zhong Wang},
  journal   = {Phys. Rev. Lett.},
  title     = {Non-Hermitian Chern Bands},
  year      = {2018},
  month     = {sep},
  number    = {13},
  pages     = {136802},
  volume    = {121},
  doi       = {10.1103/physrevlett.121.136802},
  publisher = {American Physical Society ({APS})},
}

@Article{Mostafazadeh2002,
  author    = {Ali Mostafazadeh},
  journal   = {J. Math. Phys.},
  title     = {Pseudo-Hermiticity versus {PT} symmetry: The necessary condition for the reality of the spectrum of a non-Hermitian Hamiltonian},
  year      = {2002},
  month     = {jan},
  number    = {1},
  pages     = {205--214},
  volume    = {43},
  doi       = {10.1063/1.1418246},
  publisher = {{AIP} Publishing},
}

@Article{Kawabata2021,
  author    = {Kohei Kawabata and Ken Shiozaki and Shinsei Ryu},
  journal   = {Phys. Rev. Lett.},
  title     = {Topological Field Theory of Non-Hermitian Systems},
  year      = {2021},
  month     = {may},
  number    = {21},
  pages     = {216405},
  volume    = {126},
  doi       = {10.1103/physrevlett.126.216405},
  publisher = {American Physical Society ({APS})},
}

@Article{Wojcik_2020,
  author    = {Charles C. Wojcik and Xiao-Qi Sun and Tom{\'{a}}{\v{s}} Bzdu{\v{s}}ek and Shanhui Fan},
  journal   = {Phys. Rev. B},
  title     = {Homotopy characterization of non-Hermitian Hamiltonians},
  year      = {2020},
  month     = {may},
  number    = {20},
  pages     = {205417},
  volume    = {101},
  doi       = {10.1103/physrevb.101.205417},
  publisher = {American Physical Society ({APS})},
}

@Article{Rodriguez_Nieva_2019,
  author    = {Joaquin F. Rodriguez-Nieva and Mathias S. Scheurer},
  journal   = {Nat. Phys.},
  title     = {Identifying topological order through unsupervised machine learning},
  year      = {2019},
  month     = {may},
  number    = {8},
  pages     = {790--795},
  volume    = {15},
  doi       = {10.1038/s41567-019-0512-x},
  publisher = {Springer Science and Business Media {LLC}},
}

@Article{Takata2018,
  author    = {Kenta Takata and Masaya Notomi},
  journal   = {Phys. Rev. Lett.},
  title     = {Photonic Topological Insulating Phase Induced Solely by Gain and Loss},
  year      = {2018},
  month     = {nov},
  number    = {21},
  pages     = {213902},
  volume    = {121},
  doi       = {10.1103/physrevlett.121.213902},
  publisher = {American Physical Society ({APS})},
}

@Article{Xue_2020,
  author    = {Haoran Xue and Qiang Wang and Baile Zhang and Y.{\hspace{0.167em}}D. Chong},
  journal   = {Phys. Rev. Lett.},
  title     = {Non-Hermitian Dirac Cones},
  year      = {2020},
  month     = {jun},
  number    = {23},
  pages     = {236403},
  volume    = {124},
  doi       = {10.1103/physrevlett.124.236403},
  publisher = {American Physical Society ({APS})},
}

@Article{Gao_2020,
  author    = {He Gao and Haoran Xue and Qiang Wang and Zhongming Gu and Tuo Liu and Jie Zhu and Baile Zhang},
  journal   = {Phys. Rev. B},
  title     = {Observation of topological edge states induced solely by non-Hermiticity in an acoustic crystal},
  year      = {2020},
  month     = {may},
  number    = {18},
  pages     = {180303},
  volume    = {101},
  doi       = {10.1103/physrevb.101.180303},
  publisher = {American Physical Society ({APS})},
}

@Article{Teo2010,
  author    = {Teo, Jeffrey C. Y. and Kane, C. L.},
  journal   = {Phys. Rev. B},
  title     = {Topological defects and gapless modes in insulators and superconductors},
  year      = {2010},
  issn      = {1550-235X},
  month     = sep,
  number    = {11},
  pages     = {115120},
  volume    = {82},
  doi       = {10.1103/physrevb.82.115120},
  publisher = {American Physical Society (APS)},
}

@Article{Oezdemir2019,
  author    = {Ozdemir, S. K. and Rotter, S. and Nori, F. and Yang, L.},
  journal   = {Nat. Mater.},
  title     = {Parity–time symmetry and exceptional points in photonics},
  year      = {2019},
  issn      = {1476-4660},
  month     = apr,
  number    = {8},
  pages     = {783--798},
  volume    = {18},
  doi       = {10.1038/s41563-019-0304-9},
  publisher = {Springer Science and Business Media LLC},
}

@Article{Feng2017,
  author    = {Feng, Liang and El-Ganainy, Ramy and Ge, Li},
  journal   = {Nat. Photonics},
  title     = {Non-Hermitian photonics based on parity–time symmetry},
  year      = {2017},
  issn      = {1749-4893},
  month     = nov,
  number    = {12},
  pages     = {752--762},
  volume    = {11},
  doi       = {10.1038/s41566-017-0031-1},
  publisher = {Springer Science and Business Media LLC},
}

@Article{Kawabata2019a,
  author    = {Kawabata, Kohei and Bessho, Takumi and Sato, Masatoshi},
  journal   = {Phys. Rev. Lett.},
  title     = {Classification of Exceptional Points and Non-Hermitian Topological Semimetals},
  year      = {2019},
  issn      = {1079-7114},
  month     = aug,
  number    = {6},
  pages     = {066405},
  volume    = {123},
  doi       = {10.1103/physrevlett.123.066405},
  publisher = {American Physical Society (APS)},
}

@Article{Lieu2020,
  author    = {Lieu, Simon and McGinley, Max and Cooper, Nigel R.},
  journal   = {Phys. Rev. Lett.},
  title     = {Tenfold Way for Quadratic Lindbladians},
  year      = {2020},
  issn      = {1079-7114},
  month     = jan,
  number    = {4},
  pages     = {040401},
  volume    = {124},
  doi       = {10.1103/physrevlett.124.040401},
  publisher = {American Physical Society (APS)},
}

@Article{Bender2007,
  author    = {Bender, Carl M},
  journal   = {Rep. Prog. Phys.},
  title     = {Making sense of non-Hermitian Hamiltonians},
  year      = {2007},
  issn      = {1361-6633},
  month     = may,
  number    = {6},
  pages     = {947--1018},
  volume    = {70},
  doi       = {10.1088/0034-4885/70/6/r03},
  publisher = {IOP Publishing},
}

@Article{Li2023,
  author    = {Li, Yandong and Ao, Yutian and Hu, Xiaoyong and Lu, Cuicui and Chan, C. T. and Gong, Qihuang},
  journal   = {Laser Photonics Rev.},
  title     = {Unsupervised Learning of non‐Hermitian Photonic Bulk Topology},
  year      = {2023},
  issn      = {1863-8899},
  month     = aug,
  number    = {2300481},
  pages     = {2300481},
  volume    = {17},
  doi       = {10.1002/lpor.202300481},
  publisher = {Wiley},
}

@Article{Noether1918,
  author  = {Noether, E.},
  journal = {Nachrichten von der Gesellschaft der Wissenschaften zu Gottingen, Mathematisch-Physikalische Klasse},
  title   = {Invariante Variationsprobleme},
  year    = {1918},
  pages   = {235-257},
  volume  = {1918},
}

@Article{Doppler2016,
  author    = {Doppler, Jörg and Mailybaev, Alexei A. and Böhm, Julian and Kuhl, Ulrich and Girschik, Adrian and Libisch, Florian and Milburn, Thomas J. and Rabl, Peter and Moiseyev, Nimrod and Rotter, Stefan},
  journal   = {Nature},
  title     = {Dynamically encircling an exceptional point for asymmetric mode switching},
  year      = {2016},
  issn      = {1476-4687},
  month     = jul,
  number    = {7618},
  pages     = {76--79},
  volume    = {537},
  doi       = {10.1038/nature18605},
  publisher = {Springer Science and Business Media LLC},
}

@Article{Nasari2022,
  author    = {Nasari, Hadiseh and Lopez-Galmiche, Gisela and Lopez-Aviles, Helena E. and Schumer, Alexander and Hassan, Absar U. and Zhong, Qi and Rotter, Stefan and LiKamWa, Patrick and Christodoulides, Demetrios N. and Khajavikhan, Mercedeh},
  journal   = {Nature},
  title     = {Observation of chiral state transfer without encircling an exceptional point},
  year      = {2022},
  issn      = {1476-4687},
  month     = may,
  number    = {7909},
  pages     = {256--261},
  volume    = {605},
  doi       = {10.1038/s41586-022-04542-2},
  publisher = {Springer Science and Business Media LLC},
}

@Article{Chen2017,
  author    = {Chen, Weijian and Kaya \"Ozdemir, Sahin and Zhao, Guangming and Wiersig, Jan and Yang, Lan},
  journal   = {Nature},
  title     = {Exceptional points enhance sensing in an optical microcavity},
  year      = {2017},
  issn      = {1476-4687},
  month     = aug,
  number    = {7666},
  pages     = {192--196},
  volume    = {548},
  doi       = {10.1038/nature23281},
  publisher = {Springer Science and Business Media LLC},
}

@Article{Hodaei2017,
  author    = {Hodaei, Hossein and Hassan, Absar U. and Wittek, Steffen and Garcia-Gracia, Hipolito and El-Ganainy, Ramy and Christodoulides, Demetrios N. and Khajavikhan, Mercedeh},
  journal   = {Nature},
  title     = {Enhanced sensitivity at higher-order exceptional points},
  year      = {2017},
  issn      = {1476-4687},
  month     = aug,
  number    = {7666},
  pages     = {187--191},
  volume    = {548},
  doi       = {10.1038/nature23280},
  publisher = {Springer Science and Business Media LLC},
}

@Article{Hodaei2014,
  author    = {Hodaei, Hossein and Miri, Mohammad-Ali and Heinrich, Matthias and Christodoulides, Demetrios N. and Khajavikhan, Mercedeh},
  journal   = {Science},
  title     = {Parity-time–symmetric microring lasers},
  year      = {2014},
  issn      = {1095-9203},
  month     = nov,
  number    = {6212},
  pages     = {975--978},
  volume    = {346},
  doi       = {10.1126/science.1258480},
  publisher = {American Association for the Advancement of Science (AAAS)},
}

@Article{Feng2014,
  author    = {Feng, Liang and Wong, Zi Jing and Ma, Ren-Min and Wang, Yuan and Zhang, Xiang},
  journal   = {Science},
  title     = {Single-mode laser by parity-time symmetry breaking},
  year      = {2014},
  issn      = {1095-9203},
  month     = nov,
  number    = {6212},
  pages     = {972--975},
  volume    = {346},
  doi       = {10.1126/science.1258479},
  publisher = {American Association for the Advancement of Science (AAAS)},
}

@Article{Miri2019,
  author    = {Miri, Mohammad-Ali and Alù, Andrea},
  journal   = {Science},
  title     = {Exceptional points in optics and photonics},
  year      = {2019},
  issn      = {1095-9203},
  month     = jan,
  number    = {6422},
  pages     = {eaar7709},
  volume    = {363},
  doi       = {10.1126/science.aar7709},
  publisher = {American Association for the Advancement of Science (AAAS)},
}

@Article{Li2023a,
  author    = {Li, Aodong and Wei, Heng and Cotrufo, Michele and Chen, Weijin and Mann, Sander and Ni, Xiang and Xu, Bingcong and Chen, Jianfeng and Wang, Jian and Fan, Shanhui and Qiu, Cheng-Wei and Alù, Andrea and Chen, Lin},
  journal   = {Nat. Nanotechnol.},
  title     = {Exceptional points and non-Hermitian photonics at the nanoscale},
  year      = {2023},
  issn      = {1748-3395},
  month     = jun,
  number    = {7},
  pages     = {706--720},
  volume    = {18},
  doi       = {10.1038/s41565-023-01408-0},
  publisher = {Springer Science and Business Media LLC},
}

@Article{ElGanainy2018,
  author    = {El-Ganainy, Ramy and Makris, Konstantinos G. and Khajavikhan, Mercedeh and Musslimani, Ziad H. and Rotter, Stefan and Christodoulides, Demetrios N.},
  journal   = {Nat. Phys.},
  title     = {Non-Hermitian physics and PT symmetry},
  year      = {2018},
  issn      = {1745-2481},
  month     = jan,
  number    = {1},
  pages     = {11--19},
  volume    = {14},
  doi       = {10.1038/nphys4323},
  publisher = {Springer Science and Business Media LLC},
}

@Article{Moiseyev2011,
  author    = {Moiseyev, Nimrod},
  journal   = {Cambridge University Press},
  title     = {Non-Hermitian Quantum Mechanics},
  year      = {2011},
  month     = feb,
  doi       = {10.1017/cbo9780511976186},
  isbn      = {9780511976186},
  publisher = {Cambridge University Press},
}

@Article{Gao2021,
  author    = {Gao, He and Xue, Haoran and Gu, Zhongming and Liu, Tuo and Zhu, Jie and Zhang, Baile},
  journal   = {Nat. Commun.},
  title     = {Non-Hermitian route to higher-order topology in an acoustic crystal},
  year      = {2021},
  issn      = {2041-1723},
  month     = mar,
  number    = {1},
  pages     = {1888},
  volume    = {12},
  doi       = {10.1038/s41467-021-22223-y},
  publisher = {Springer Science and Business Media LLC},
}

@Article{Ding2022,
  author    = {Ding, Kun and Fang, Chen and Ma, Guancong},
  journal   = {Nat. Rev. Phys.},
  title     = {Non-Hermitian topology and exceptional-point geometries},
  year      = {2022},
  issn      = {2522-5820},
  month     = oct,
  number    = {12},
  pages     = {745--760},
  volume    = {4},
  doi       = {10.1038/s42254-022-00516-5},
  publisher = {Springer Science and Business Media LLC},
}

@Article{Coulais2020,
  author    = {Coulais, Corentin and Fleury, Romain and van Wezel, Jasper},
  journal   = {Nat. Phys.},
  title     = {Topology and broken Hermiticity},
  year      = {2020},
  issn      = {1745-2481},
  month     = nov,
  number    = {1},
  pages     = {9--13},
  volume    = {17},
  doi       = {10.1038/s41567-020-01093-z},
  publisher = {Springer Science and Business Media LLC},
}

@Article{Hu2021,
  author    = {Hu, Bolun and Zhang, Zhiwang and Zhang, Haixiao and Zheng, Liyang and Xiong, Wei and Yue, Zichong and Wang, Xiaoyu and Xu, Jianyi and Cheng, Ying and Liu, Xiaojun and Christensen, Johan},
  journal   = {Nature},
  title     = {Non-Hermitian topological whispering gallery},
  year      = {2021},
  issn      = {1476-4687},
  month     = sep,
  number    = {7878},
  pages     = {655--659},
  volume    = {597},
  doi       = {10.1038/s41586-021-03833-4},
  publisher = {Springer Science and Business Media LLC},
}

@Article{Wang2022,
  author    = {Wang, Wei and Wang, Xulong and Ma, Guancong},
  journal   = {Nature},
  title     = {Non-Hermitian morphing of topological modes},
  year      = {2022},
  issn      = {1476-4687},
  month     = aug,
  number    = {7921},
  pages     = {50--55},
  volume    = {608},
  doi       = {10.1038/s41586-022-04929-1},
  publisher = {Springer Science and Business Media LLC},
}

@Article{Wang2016,
  author    = {Wang, Lei},
  journal   = {Phys. Rev. B},
  title     = {Discovering phase transitions with unsupervised learning},
  year      = {2016},
  issn      = {2469-9969},
  month     = nov,
  number    = {19},
  pages     = {195105},
  volume    = {94},
  doi       = {10.1103/physrevb.94.195105},
  publisher = {American Physical Society (APS)},
}

@Article{Wetzel2017,
  author    = {Wetzel, Sebastian J.},
  journal   = {Phys. Rev. E},
  title     = {Unsupervised learning of phase transitions: From principal component analysis to variational autoencoders},
  year      = {2017},
  issn      = {2470-0053},
  month     = aug,
  number    = {2},
  pages     = {022140},
  volume    = {96},
  doi       = {10.1103/physreve.96.022140},
  publisher = {American Physical Society (APS)},
}

@Article{Long2020,
  author    = {Long, Yang and Ren, Jie and Chen, Hong},
  journal   = {Phys. Rev. Lett.},
  title     = {Unsupervised Manifold Clustering of Topological Phononics},
  year      = {2020},
  issn      = {1079-7114},
  month     = may,
  number    = {18},
  pages     = {185501},
  volume    = {124},
  doi       = {10.1103/physrevlett.124.185501},
  publisher = {American Physical Society (APS)},
}

@Article{Che2020,
  author    = {Che, Yanming and Gneiting, Clemens and Liu, Tao and Nori, Franco},
  journal   = {Phys. Rev. B},
  title     = {Topological quantum phase transitions retrieved through unsupervised machine learning},
  year      = {2020},
  issn      = {2469-9969},
  month     = oct,
  number    = {13},
  pages     = {134213},
  volume    = {102},
  doi       = {10.1103/physrevb.102.134213},
  publisher = {American Physical Society (APS)},
}

@Article{Balabanov2020,
  author    = {Balabanov, Oleksandr and Granath, Mats},
  journal   = {Phys. Rev. Res.},
  title     = {Unsupervised learning using topological data augmentation},
  year      = {2020},
  issn      = {2643-1564},
  month     = mar,
  number    = {1},
  pages     = {013354},
  volume    = {2},
  doi       = {10.1103/physrevresearch.2.013354},
  publisher = {American Physical Society (APS)},
}

@Article{Park2022,
  author    = {Park, Sungjoon and Hwang, Yoonseok and Yang, Bohm-Jung},
  journal   = {Phys. Rev. B},
  title     = {Unsupervised learning of topological phase diagram using topological data analysis},
  year      = {2022},
  issn      = {2469-9969},
  month     = may,
  number    = {19},
  pages     = {195115},
  volume    = {105},
  doi       = {10.1103/physrevb.105.195115},
  publisher = {American Physical Society (APS)},
}

@Article{Ma2022,
  author    = {Ma, Nannan and Gong, Jiangbin},
  journal   = {Phys. Rev. Res.},
  title     = {Unsupervised identification of Floquet topological phase boundaries},
  year      = {2022},
  issn      = {2643-1564},
  month     = mar,
  number    = {1},
  pages     = {013234},
  volume    = {4},
  doi       = {10.1103/physrevresearch.4.013234},
  publisher = {American Physical Society (APS)},
}

@Article{Luo2019,
  author    = {Luo, Xi-Wang and Zhang, Chuanwei},
  journal   = {Phys. Rev. Lett.},
  title     = {Higher-Order Topological Corner States Induced by Gain and Loss},
  year      = {2019},
  issn      = {1079-7114},
  month     = aug,
  number    = {7},
  pages     = {073601},
  volume    = {123},
  doi       = {10.1103/physrevlett.123.073601},
  publisher = {American Physical Society (APS)},
}

@Article{Rueter2010,
  author    = {Rüter, Christian E. and Makris, Konstantinos G. and El-Ganainy, Ramy and Christodoulides, Demetrios N. and Segev, Mordechai and Kip, Detlef},
  journal   = {Nat. Phys.},
  title     = {Observation of parity–time symmetry in optics},
  year      = {2010},
  issn      = {1745-2481},
  month     = jan,
  number    = {3},
  pages     = {192--195},
  volume    = {6},
  doi       = {10.1038/nphys1515},
  publisher = {Springer Science and Business Media LLC},
}

@Article{Okuma2020,
  author    = {Okuma, Nobuyuki and Kawabata, Kohei and Shiozaki, Ken and Sato, Masatoshi},
  journal   = {Phys. Rev. Lett.},
  title     = {Topological Origin of Non-Hermitian Skin Effects},
  year      = {2020},
  issn      = {1079-7114},
  month     = feb,
  number    = {8},
  pages     = {086801},
  volume    = {124},
  doi       = {10.1103/physrevlett.124.086801},
  publisher = {American Physical Society (APS)},
}

@Article{McDonald2023,
  author    = {McDonald, Alexander and Clerk, Aashish A.},
  journal   = {Phys. Rev. Res.},
  title     = {Third quantization of open quantum systems: Dissipative symmetries and connections to phase-space and Keldysh field-theory formulations},
  year      = {2023},
  issn      = {2643-1564},
  month     = aug,
  number    = {3},
  pages     = {033107},
  volume    = {5},
  doi       = {10.1103/physrevresearch.5.033107},
  publisher = {American Physical Society (APS)},
}

@Article{Zhu2022,
  author    = {Zhu, Bofeng and Wang, Qiang and Leykam, Daniel and Xue, Haoran and Wang, Qi Jie and Chong, Y. D.},
  journal   = {Phys. Rev. Lett.},
  title     = {Anomalous Single-Mode Lasing Induced by Nonlinearity and the Non-Hermitian Skin Effect},
  year      = {2022},
  issn      = {1079-7114},
  month     = jun,
  number    = {1},
  pages     = {013903},
  volume    = {129},
  doi       = {10.1103/physrevlett.129.013903},
  publisher = {American Physical Society (APS)},
}

@Article{Liu2023,
  author    = {Liu, Dongjue and Wang, Zihao and Cheng, Zheyu and Hu, Hao and Wang, Qijie and Xue, Haoran and Zhang, Baile and Luo, Yu},
  journal   = {Laser Photonics Rev.},
  title     = {Simultaneous Manipulation of Line‐Gap and Point‐Gap Topologies in Non‐Hermitian Lattices},
  year      = {2023},
  issn      = {1863-8899},
  month     = feb,
  number    = {4},
  pages     = {2200371},
  volume    = {17},
  doi       = {10.1002/lpor.202200371},
  publisher = {Wiley},
}

@Article{Harari2018,
  author    = {Harari, Gal and Bandres, Miguel A. and Lumer, Yaakov and Rechtsman, Mikael C. and Chong, Y. D. and Khajavikhan, Mercedeh and Christodoulides, Demetrios N. and Segev, Mordechai},
  journal   = {Science},
  title     = {Topological insulator laser: Theory},
  year      = {2018},
  issn      = {1095-9203},
  month     = mar,
  number    = {6381},
  pages     = {eaar4003},
  volume    = {359},
  doi       = {10.1126/science.aar4003},
  publisher = {American Association for the Advancement of Science (AAAS)},
}

@Article{Bandres2018,
  author    = {Bandres, Miguel A. and Wittek, Steffen and Harari, Gal and Parto, Midya and Ren, Jinhan and Segev, Mordechai and Christodoulides, Demetrios N. and Khajavikhan, Mercedeh},
  journal   = {Science},
  title     = {Topological insulator laser: Experiments},
  year      = {2018},
  issn      = {1095-9203},
  month     = mar,
  number    = {6381},
  pages     = {eaar4005},
  volume    = {359},
  doi       = {10.1126/science.aar4005},
  publisher = {American Association for the Advancement of Science (AAAS)},
}

@Article{Zeng2020,
  author    = {Zeng, Yongquan and Chattopadhyay, Udvas and Zhu, Bofeng and Qiang, Bo and Li, Jinghao and Jin, Yuhao and Li, Lianhe and Davies, Alexander Giles and Linfield, Edmund Harold and Zhang, Baile and Chong, Yidong and Wang, Qi Jie},
  journal   = {Nature},
  title     = {Electrically pumped topological laser with valley edge modes},
  year      = {2020},
  issn      = {1476-4687},
  month     = feb,
  number    = {7794},
  pages     = {246--250},
  volume    = {578},
  doi       = {10.1038/s41586-020-1981-x},
  publisher = {Springer Science and Business Media LLC},
}

@Article{Amelio2020,
  author    = {Amelio, Ivan and Carusotto, Iacopo},
  journal   = {Phys. Rev. X},
  title     = {Theory of the Coherence of Topological Lasers},
  year      = {2020},
  issn      = {2160-3308},
  month     = dec,
  number    = {4},
  pages     = {041060},
  volume    = {10},
  doi       = {10.1103/physrevx.10.041060},
  publisher = {American Physical Society (APS)},
}

@Article{Shao2019,
  author    = {Shao, Zeng-Kai and Chen, Hua-Zhou and Wang, Suo and Mao, Xin-Rui and Yang, Zhen-Qian and Wang, Shao-Lei and Wang, Xing-Xiang and Hu, Xiao and Ma, Ren-Min},
  journal   = {Nat. Nanotechnol.},
  title     = {A high-performance topological bulk laser based on band-inversion-induced reflection},
  year      = {2019},
  issn      = {1748-3395},
  month     = dec,
  number    = {1},
  pages     = {67--72},
  volume    = {15},
  doi       = {10.1038/s41565-019-0584-x},
  publisher = {Springer Science and Business Media LLC},
}

@Article{Borgnia2020,
  author    = {Borgnia, Dan S. and Kruchkov, Alex Jura and Slager, Robert-Jan},
  journal   = {Phys. Rev. Lett.},
  title     = {Non-Hermitian Boundary Modes and Topology},
  year      = {2020},
  issn      = {1079-7114},
  month     = feb,
  number    = {5},
  pages     = {056802},
  volume    = {124},
  doi       = {10.1103/physrevlett.124.056802},
  publisher = {American Physical Society (APS)},
}

@Article{Carrasquilla2017,
  author    = {Carrasquilla, Juan and Melko, Roger G.},
  journal   = {Nat. Phys.},
  title     = {Machine learning phases of matter},
  year      = {2017},
  issn      = {1745-2481},
  month     = feb,
  number    = {5},
  pages     = {431--434},
  volume    = {13},
  doi       = {10.1038/nphys4035},
  publisher = {Springer Science and Business Media LLC},
}

@Article{Nieuwenburg2017,
  author    = {van Nieuwenburg, Evert P. L. and Liu, Ye-Hua and Huber, Sebastian D.},
  journal   = {Nat. Phys.},
  title     = {Learning phase transitions by confusion},
  year      = {2017},
  issn      = {1745-2481},
  month     = feb,
  number    = {5},
  pages     = {435--439},
  volume    = {13},
  doi       = {10.1038/nphys4037},
  publisher = {Springer Science and Business Media LLC},
}

@Article{Qi2011,
  author    = {Qi, Xiao-Liang and Zhang, Shou-Cheng},
  journal   = {Rev. Mod. Phys.},
  title     = {Topological insulators and superconductors},
  year      = {2011},
  issn      = {1539-0756},
  month     = oct,
  number    = {4},
  pages     = {1057--1110},
  volume    = {83},
  doi       = {10.1103/revmodphys.83.1057},
  publisher = {American Physical Society (APS)},
}

@Article{Hasan2010,
  author    = {Hasan, M. Z. and Kane, C. L.},
  journal   = {Rev. Mod. Phys.},
  title     = {Colloquium: Topological insulators},
  year      = {2010},
  issn      = {1539-0756},
  month     = nov,
  number    = {4},
  pages     = {3045--3067},
  volume    = {82},
  doi       = {10.1103/revmodphys.82.3045},
  publisher = {American Physical Society (APS)},
}

@Article{Haldane2017,
  author    = {Haldane, F. Duncan M.},
  journal   = {Rev. Mod. Phys.},
  title     = {Nobel Lecture: Topological quantum matter},
  year      = {2017},
  issn      = {1539-0756},
  month     = oct,
  number    = {4},
  pages     = {040502},
  volume    = {89},
  doi       = {10.1103/revmodphys.89.040502},
  publisher = {American Physical Society (APS)},
}

@Article{Bansil2016,
  author    = {Bansil, A. and Lin, Hsin and Das, Tanmoy},
  journal   = {Rev. Mod. Phys.},
  title     = {Colloquium: Topological band theory},
  year      = {2016},
  issn      = {1539-0756},
  month     = jun,
  number    = {2},
  pages     = {021004},
  volume    = {88},
  doi       = {10.1103/revmodphys.88.021004},
  publisher = {American Physical Society (APS)},
}

@Article{Wang2023,
  author    = {Wang, Hanchen and Fu, Tianfan and Du, Yuanqi and Gao, Wenhao and Huang, Kexin and Liu, Ziming and Chandak, Payal and Liu, Shengchao and Van Katwyk, Peter and Deac, Andreea and Anandkumar, Anima and Bergen, Karianne and Gomes, Carla P. and Ho, Shirley and Kohli, Pushmeet and Lasenby, Joan and Leskovec, Jure and Liu, Tie-Yan and Manrai, Arjun and Marks, Debora and Ramsundar, Bharath and Song, Le and Sun, Jimeng and Tang, Jian and Veličković, Petar and Welling, Max and Zhang, Linfeng and Coley, Connor W. and Bengio, Yoshua and Zitnik, Marinka},
  journal   = {Nature},
  title     = {Scientific discovery in the age of artificial intelligence},
  year      = {2023},
  issn      = {1476-4687},
  month     = aug,
  number    = {7972},
  pages     = {47--60},
  volume    = {620},
  doi       = {10.1038/s41586-023-06221-2},
  publisher = {Springer Science and Business Media LLC},
}

@Article{Krenn2022,
  author    = {Krenn, Mario and Pollice, Robert and Guo, Si Yue and Aldeghi, Matteo and Cervera-Lierta, Alba and Friederich, Pascal and dos Passos Gomes, Gabriel and Häse, Florian and Jinich, Adrian and Nigam, AkshatKumar and Yao, Zhenpeng and Aspuru-Guzik, Alán},
  journal   = {Nat. Rev. Phys.},
  title     = {On scientific understanding with artificial intelligence},
  year      = {2022},
  issn      = {2522-5820},
  month     = oct,
  number    = {12},
  pages     = {761--769},
  volume    = {4},
  doi       = {10.1038/s42254-022-00518-3},
  publisher = {Springer Science and Business Media LLC},
}

@Article{Mehta2019,
  author    = {Mehta, Pankaj and Bukov, Marin and Wang, Ching-Hao and Day, Alexandre G.R. and Richardson, Clint and Fisher, Charles K. and Schwab, David J.},
  journal   = {Phys. Rep.},
  title     = {A high-bias, low-variance introduction to Machine Learning for physicists},
  year      = {2019},
  issn      = {0370-1573},
  month     = may,
  pages     = {1--124},
  volume    = {810},
  doi       = {10.1016/j.physrep.2019.03.001},
  publisher = {Elsevier BV},
}

@Article{Dunjko2018,
  author    = {Dunjko, Vedran and Briegel, Hans J},
  journal   = {Rep. Prog. Phys.},
  title     = {Machine learning $\&$ artificial intelligence in the quantum domain: a review of recent progress},
  year      = {2018},
  issn      = {1361-6633},
  month     = jun,
  number    = {7},
  pages     = {074001},
  volume    = {81},
  doi       = {10.1088/1361-6633/aab406},
  publisher = {IOP Publishing},
}

@Article{Carleo2019,
  author    = {Carleo, Giuseppe and Cirac, Ignacio and Cranmer, Kyle and Daudet, Laurent and Schuld, Maria and Tishby, Naftali and Vogt-Maranto, Leslie and Zdeborová, Lenka},
  journal   = {Rev. Mod. Phys.},
  title     = {Machine learning and the physical sciences},
  year      = {2019},
  issn      = {1539-0756},
  month     = dec,
  number    = {4},
  pages     = {045002},
  volume    = {91},
  doi       = {10.1103/revmodphys.91.045002},
  publisher = {American Physical Society (APS)},
}

@Article{Carleo2017,
  author    = {Carleo, Giuseppe and Troyer, Matthias},
  journal   = {Science},
  title     = {Solving the quantum many-body problem with artificial neural networks},
  year      = {2017},
  issn      = {1095-9203},
  month     = feb,
  number    = {6325},
  pages     = {602--606},
  volume    = {355},
  doi       = {10.1126/science.aag2302},
  publisher = {American Association for the Advancement of Science (AAAS)},
}

@Article{Hinton2006,
  author    = {Hinton, G. E. and Salakhutdinov, R. R.},
  journal   = {Science},
  title     = {Reducing the Dimensionality of Data with Neural Networks},
  year      = {2006},
  issn      = {1095-9203},
  month     = jul,
  number    = {5786},
  pages     = {504--507},
  volume    = {313},
  doi       = {10.1126/science.1127647},
  publisher = {American Association for the Advancement of Science (AAAS)},
}

@Article{Melko2019,
  author    = {Melko, Roger G. and Carleo, Giuseppe and Carrasquilla, Juan and Cirac, J. Ignacio},
  journal   = {Nat. Phys.},
  title     = {Restricted Boltzmann machines in quantum physics},
  year      = {2019},
  issn      = {1745-2481},
  month     = jun,
  number    = {9},
  pages     = {887--892},
  volume    = {15},
  doi       = {10.1038/s41567-019-0545-1},
  publisher = {Springer Science and Business Media LLC},
}

@Article{Mnih2015,
  author    = {Mnih, Volodymyr and Kavukcuoglu, Koray and Silver, David and Rusu, Andrei A. and Veness, Joel and Bellemare, Marc G. and Graves, Alex and Riedmiller, Martin and Fidjeland, Andreas K. and Ostrovski, Georg and Petersen, Stig and Beattie, Charles and Sadik, Amir and Antonoglou, Ioannis and King, Helen and Kumaran, Dharshan and Wierstra, Daan and Legg, Shane and Hassabis, Demis},
  journal   = {Nature},
  title     = {Human-level control through deep reinforcement learning},
  year      = {2015},
  issn      = {1476-4687},
  month     = feb,
  number    = {7540},
  pages     = {529--533},
  volume    = {518},
  doi       = {10.1038/nature14236},
  publisher = {Springer Science and Business Media LLC},
}

@Article{Silver2016,
  author    = {Silver, David and Huang, Aja and Maddison, Chris J. and Guez, Arthur and Sifre, Laurent and van den Driessche, George and Schrittwieser, Julian and Antonoglou, Ioannis and Panneershelvam, Veda and Lanctot, Marc and Dieleman, Sander and Grewe, Dominik and Nham, John and Kalchbrenner, Nal and Sutskever, Ilya and Lillicrap, Timothy and Leach, Madeleine and Kavukcuoglu, Koray and Graepel, Thore and Hassabis, Demis},
  journal   = {Nature},
  title     = {Mastering the game of Go with deep neural networks and tree search},
  year      = {2016},
  issn      = {1476-4687},
  month     = jan,
  number    = {7587},
  pages     = {484--489},
  volume    = {529},
  doi       = {10.1038/nature16961},
  publisher = {Springer Science and Business Media LLC},
}

@Article{Silver2017,
  author    = {Silver, David and Schrittwieser, Julian and Simonyan, Karen and Antonoglou, Ioannis and Huang, Aja and Guez, Arthur and Hubert, Thomas and Baker, Lucas and Lai, Matthew and Bolton, Adrian and Chen, Yutian and Lillicrap, Timothy and Hui, Fan and Sifre, Laurent and van den Driessche, George and Graepel, Thore and Hassabis, Demis},
  journal   = {Nature},
  title     = {Mastering the game of Go without human knowledge},
  year      = {2017},
  issn      = {1476-4687},
  month     = oct,
  number    = {7676},
  pages     = {354--359},
  volume    = {550},
  doi       = {10.1038/nature24270},
  publisher = {Springer Science and Business Media LLC},
}

@Article{Silver2018,
  author    = {Silver, David and Hubert, Thomas and Schrittwieser, Julian and Antonoglou, Ioannis and Lai, Matthew and Guez, Arthur and Lanctot, Marc and Sifre, Laurent and Kumaran, Dharshan and Graepel, Thore and Lillicrap, Timothy and Simonyan, Karen and Hassabis, Demis},
  journal   = {Science},
  title     = {A general reinforcement learning algorithm that masters chess, shogi, and Go through self-play},
  year      = {2018},
  issn      = {1095-9203},
  month     = dec,
  number    = {6419},
  pages     = {1140--1144},
  volume    = {362},
  doi       = {10.1126/science.aar6404},
  publisher = {American Association for the Advancement of Science (AAAS)},
}

@Article{Senior2020,
  author    = {Senior, Andrew W. and Evans, Richard and Jumper, John and Kirkpatrick, James and Sifre, Laurent and Green, Tim and Qin, Chongli and Žídek, Augustin and Nelson, Alexander W. R. and Bridgland, Alex and Penedones, Hugo and Petersen, Stig and Simonyan, Karen and Crossan, Steve and Kohli, Pushmeet and Jones, David T. and Silver, David and Kavukcuoglu, Koray and Hassabis, Demis},
  journal   = {Nature},
  title     = {Improved protein structure prediction using potentials from deep learning},
  year      = {2020},
  issn      = {1476-4687},
  month     = jan,
  number    = {7792},
  pages     = {706--710},
  volume    = {577},
  doi       = {10.1038/s41586-019-1923-7},
  publisher = {Springer Science and Business Media LLC},
}

@Article{Jumper2021,
  author    = {Jumper, John and Evans, Richard and Pritzel, Alexander and Green, Tim and Figurnov, Michael and Ronneberger, Olaf and Tunyasuvunakool, Kathryn and Bates, Russ and Žídek, Augustin and Potapenko, Anna and Bridgland, Alex and Meyer, Clemens and Kohl, Simon A. A. and Ballard, Andrew J. and Cowie, Andrew and Romera-Paredes, Bernardino and Nikolov, Stanislav and Jain, Rishub and Adler, Jonas and Back, Trevor and Petersen, Stig and Reiman, David and Clancy, Ellen and Zielinski, Michal and Steinegger, Martin and Pacholska, Michalina and Berghammer, Tamas and Bodenstein, Sebastian and Silver, David and Vinyals, Oriol and Senior, Andrew W. and Kavukcuoglu, Koray and Kohli, Pushmeet and Hassabis, Demis},
  journal   = {Nature},
  title     = {Highly accurate protein structure prediction with AlphaFold},
  year      = {2021},
  issn      = {1476-4687},
  month     = jul,
  number    = {7873},
  pages     = {583--589},
  volume    = {596},
  doi       = {10.1038/s41586-021-03819-2},
  publisher = {Springer Science and Business Media LLC},
}

@Article{Zhang2013,
  author    = {Zhang, Fan and MacDonald, Allan H. and Mele, Eugene J.},
  journal   = {Proc. Natl. Acad. Sci. U.S.A.},
  title     = {Valley Chern numbers and boundary modes in gapped bilayer graphene},
  year      = {2013},
  issn      = {1091-6490},
  month     = jun,
  number    = {26},
  pages     = {10546--10551},
  volume    = {110},
  doi       = {10.1073/pnas.1308853110},
  publisher = {Proceedings of the National Academy of Sciences},
}

@Article{Benalcazar2017,
  author    = {Benalcazar, Wladimir A. and Bernevig, B. Andrei and Hughes, Taylor L.},
  journal   = {Science},
  title     = {Quantized electric multipole insulators},
  year      = {2017},
  issn      = {1095-9203},
  month     = jul,
  number    = {6346},
  pages     = {61--66},
  volume    = {357},
  doi       = {10.1126/science.aah6442},
  publisher = {American Association for the Advancement of Science (AAAS)},
}

@Article{Liu2017,
  author    = {Liu, Feng and Wakabayashi, Katsunori},
  journal   = {Phys. Rev. Lett.},
  title     = {Novel Topological Phase with a Zero Berry Curvature},
  year      = {2017},
  issn      = {1079-7114},
  month     = feb,
  number    = {7},
  pages     = {076803},
  volume    = {118},
  doi       = {10.1103/physrevlett.118.076803},
  publisher = {American Physical Society (APS)},
}

@Article{Long2024,
  author    = {Long, Yang and Xue, Haoran and Zhang, Baile},
  journal   = {Nat. Mach. Intell.},
  title     = {Unsupervised learning of topological non-Abelian braiding in non-Hermitian bands},
  year      = {2024},
  issn      = {2522-5839},
  month     = jul,
  number    = {8},
  pages     = {904--910},
  volume    = {6},
  doi       = {10.1038/s42256-024-00871-1},
  publisher = {Springer Science and Business Media LLC},
}

@Article{Chen2024,
  author    = {Chen, Jiangzhi and Wang, Zi and Tan, Yu-Tao and Wang, Ce and Ren, Jie},
  journal   = {Commun. Phys.},
  title     = {Machine learning of knot topology in non-Hermitian band braids},
  year      = {2024},
  issn      = {2399-3650},
  month     = jun,
  number    = {1},
  pages     = {209},
  volume    = {7},
  doi       = {10.1038/s42005-024-01710-w},
  publisher = {Springer Science and Business Media LLC},
}

@Article{Yu2022,
  author    = {Yu, Yefei and Yu, Li-Wei and Zhang, Wengang and Zhang, Huili and Ouyang, Xiaolong and Liu, Yanqing and Deng, Dong-Ling and Duan, L.-M.},
  journal   = {npj Quantum Information},
  title     = {Experimental unsupervised learning of non-Hermitian knotted phases with solid-state spins},
  year      = {2022},
  issn      = {2056-6387},
  month     = sep,
  number    = {1},
  pages     = {116},
  volume    = {8},
  doi       = {10.1038/s41534-022-00629-w},
  publisher = {Springer Science and Business Media LLC},
}

@Article{Kunst2018,
  author    = {Kunst, Flore K. and Edvardsson, Elisabet and Budich, Jan Carl and Bergholtz, Emil J.},
  journal   = {Phys. Rev. Lett.},
  title     = {Biorthogonal Bulk-Boundary Correspondence in Non-Hermitian Systems},
  year      = {2018},
  issn      = {1079-7114},
  month     = jul,
  number    = {2},
  pages     = {026808},
  volume    = {121},
  doi       = {10.1103/physrevlett.121.026808},
  publisher = {American Physical Society (APS)},
}

@Misc{long_github_2024_non_Hermitian,
  author       = {Yang Long},
  howpublished = {\url{https://github.com/longyangking/ml\_topological\_classification\_non\_hermitian}},
  title        = {GitHub repository},
  year         = {2024},
  journal      = {GitHub repository},
  publisher    = {GitHub},
  url          = {https://github.com/longyangking/ml\_topological\_classification\_non\_hermitian},
}

@Article{Wang2025,
  author    = {Wang, Zhiyuan and Hazzard, Kaden R. A.},
  journal   = {Nature},
  title     = {Particle exchange statistics beyond fermions and bosons},
  year      = {2025},
  issn      = {1476-4687},
  month     = jan,
  number    = {8045},
  pages     = {314--318},
  volume    = {637},
  doi       = {10.1038/s41586-024-08262-7},
  publisher = {Springer Science and Business Media LLC},
}

@Article{Kitaev2001,
  author    = {Kitaev, A Yu},
  journal   = {Physics-Uspekhi},
  title     = {Unpaired Majorana fermions in quantum wires},
  year      = {2001},
  issn      = {1468-4780},
  month     = oct,
  number    = {10S},
  pages     = {131--136},
  volume    = {44},
  doi       = {10.1070/1063-7869/44/10s/s29},
  publisher = {Uspekhi Fizicheskikh Nauk (UFN) Journal},
}

\end{document}